\documentclass[a4paper, 12pt]{article}

\usepackage{PRIMEarxiv}

\usepackage[utf8]{inputenc} 
\usepackage[T1]{fontenc}    
\usepackage{hyperref}       
\usepackage{url}            
\usepackage{booktabs}       
\usepackage{amsfonts}       
\usepackage{nicefrac}       
\usepackage{microtype}      
\usepackage{lipsum}
\usepackage{fancyhdr}       
\usepackage{graphicx}       
\graphicspath{{media/}}     

\usepackage{algorithm}
\usepackage{algpseudocode}
\usepackage{pdflscape}
\usepackage{tabularx}
\usepackage{changepage}
\usepackage{multirow}
\usepackage{amsmath}
\usepackage{xcolor, colortbl}
\usepackage{tocbibind}
\usepackage[toc,page]{appendix}
\usepackage{makecell}

\title{Invoice Discounting using Kelly Criterion by Automated Market Makers-like Implementations
}

\author{
  Peplluis R Esteva \\
  Byppay Global SL, Girona, Spain \\
  Centre of Blockchain Technology Fellow - University College London (UCL - CBT), London, UK \\
  \texttt{peplluis@byppay.com} \\
   \And
  Alberto Ballesteros Rodríguez \\
  University of Alcalá, Alcalá de Henares, Spain \\
  University of Girona, Girona, Spain \\
  Computing and Artificial Intelligence Laboratory (CAILab), Camilo José Cela University, Madrid, Spain \\
  \texttt{alberto.ballesterosr@uah.es} \\
}

\begin{document}
\maketitle

\begin{abstract}

{There is a persistent lack of funding, especially for SMEs, that cyclically worsens. The factoring and
invoice discounting market appears to address delays in paying commercial invoices: sellers bring
still-to-be-paid invoices to financial organizations, intermediaries, typically banks that provide an
advance payment. This article contains research on novel decentralized approaches to said lending
services without intermediaries by using liquidity pools and its associated heuristics, creating an
Automated Market Maker. In our approach, the contributed collateral and the invoice trades with
risk is measured with a formula: The Kelly criterion is used to calculate the optimal premium to be
contributed to a liquidity pool in the funding of the said invoices. The behavior of the algorithm is
studied in several scenarios of streams of invoices with representative amounts, collaterals, payment
delays, and nonpayments rates or mora. We completed the study with hack scenarios with bogus,
nonpayable invoices. As a result, we have created a resilient solution that performs the best with
partially collateralized invoices. The outcome is decentralized market developed with the Kelly
criterion that is reasonably resilient to a wide variety of the invoicing cases that provides sound profit
to liquidity providers, and several premium distribution policies were checked that contributed with extra resilience to the performance of the algorithm.}

\end{abstract}


\section{Introduction}

The market of invoice discounting is a market with a double‐digit potential growth rate over the next years in Europe and worldwide \cite{fabrizio2019invoice}. Invoice discounting already represents 10\% of banks’ provided credit, and it has become a major source of working capital finance globally after the restriction of bank financing due to the 2011 credit crunch \cite{wehinger2014smes}. It seems it will become even greater in the next crisis announced for 2023 an on. In particular, as an advantage, invoice discounting helps companies, especially Small and Medium Enterprises (SMEs), that have cash flow problems because of late payments from customers (i.e., invoices are usually paid in 30–90 days). For European SMEs it has surpassed loans and other forms of financing over the past decades \cite{wehinger2014smes}. Another advantage of invoice discounting is the confidentiality: namely, suppliers control the sales ledger by collecting payments as usual and sending out reminders. The customer (debtor) is not involved in the discounting process, hence it is not informed that the supplier (creditor) is getting his/her credit financed. From the point of view of the suppliers’ companies, the confidentiality is an advantage since, for easing negotiations with partners, they might not want to disclose their use of working capital finance.

As \cite{b52d2f6c50eb46a68284c83b05a1ea4d} claims, the majority of small companies that belong to this financial market are poorly served despite the fact that it is recognized that this market has the potential to address the liquidity needs of small companies. Two critical features of an invoice market are extremely difficult to evaluate for factors: authenticated identities of parties involved (seller and buyer) and whether or not an invoice has already been factored. As a result, factors can be victim of frauds by sellers that factor non-existent invoices or factor an invoice to multiple factors (double pledging). At the same time, buyers might be contacted by fraudulent factors that claims to have factored an invoice of a legitimate seller and demand payments for such invoices. Even in absence of frauds, paying the invoice to wrong organization creates a number of issues.

A quick solution would have been a centralized system where all invoices could be stored and checked by interested parties. Yet, despite of the existence of factoring markets for centuries, no such entities emerged. The main reason for such an absence is the quest for confidentiality. For different reasons, buyers, factors, and sellers have no interest of sharing information on past invoices as its disclosure can go against them.

In the scheme of centralized systems, the work presented in \cite{b52d2f6c50eb46a68284c83b05a1ea4d} proposes a third trusted party to fix all that. As \cite{mohammadzadeh2021invoice} claimed in their work, the main property that a factoring system needs to fulfill is to prevent an invoice from being factored twice. 

Blockchain can be used to remake a wide range of finance processes: inter company transactions (when there are multiple ERPs), procure-to-pay, order-to-cash, rebates, warranties, and financing (such as trade finance, letters of credit, and invoice factoring). Any place paper piles up present an opportunity for blockchain to move in and knock it down.

\subsection{Invoice discounting and factoring}

The main benefit of invoice discounting is the acceleration of cash flow from customers to suppliers: suppliers get advance payments from the bank rather than waiting for the customers to pay. Hence, thanks to the quick availability of capital, businesses can invest in expansion and growth. More specifically, one of the most relevant problems today is how to provide better and faster invoice discounting services while preventing double spending and maintaining risk low. The blockchain frameworks have the potential to provide the right solution and thus to revolutionize the invoice discounting process. The benefits for suppliers, customers and financial institutions are related to the increased transparency added to the whole discounting process and the following risk reduction for the banks due to the capability to enhance the entire process and to reduce the double spending.

About Factoring, it is a financial transaction and a type of debtor finance in which a business sells its accounts receivable (i.e., invoices) to a third party (called a factor) at a discount. A business will sometimes factor its receivable assets to meet its present and immediate cash needs. The actual problems of factoring are those of inter-mediation and those of the general area of trade finance and supply chain.

This paper works deeper in novel alternative approaches to invoice discounting or factoring where we will radically decentralise the way invoices are used. We plan to find solutions other than dumping all invoices on a ledger so that algorithms implemented as dApps or smart contracts make scoring of companies, decentralise crowd funding instead of banks. We have conceived with byppay a new solution to invoice compensation with debt soothing effects with a radically decentralised approach other than invoice uploading so that only those invoices related to the clients and suppliers of an actor are declared on a ledger with a full mechanism of PoE (Proof of Existence), and the inherent benefit of such a decentralised approach to invoice compensation is bolstered \cite{ioannou2022blockchain}.

\subsection{State of the Art of Blockchain for Funding}

For the discounter the benefits of decentralization and blockchain adoption are as follows \cite{Deloitte2022}: (i) an immutable and time stamped record of the existence of every invoice emitted by a company, (ii) an immutable and time stamped record of the debtor’s receipt, and (iii) the confirmation and verification of the invoice (against which a discounter would fund). Hence, the overall invoice discounting process will be enhanced. Indeed, the trust and security mechanisms of the blockchain allow for the elimination of on-site audits of receivables and debtors, of receivables’ notification and debtors’ verification, and of month-end reconciliation processes. Moreover, the adoption of blockchain will also allow for a fast and cheaper value transfer, in particular for cross-border payments. More specifically, the debtor’s verification of the invoice validity and of the reception of goods and services reduces significantly the risk of dispute and non-payment of that invoice. Moreover, the debtor could have an incentive to acknowledge and confirm its invoices without delay, as his/her own track record of confirming invoices would be visible on the blockchain to his/her suppliers and thus it could be used to influence the payment terms and the offered contract prices. This immutable debtor’s verification could also potentially eliminate the risk of invoice fraud for a discounter as there would be no “consensus” met for double invoicing transactions. In fact, time-stamping an invoice has a legal value: if a company attempts to assign its invoice more than once, it would prevent any subsequent assignee being a $bona fide$ purchaser for value without notice, thus protecting the first assignee.

\subsection{Proof of Existence}

In the paper \cite{fabrizio2019invoice}, they had provided evidence that the invoice discounting service might be improved by adopting approaches based on a distributed ledger. In our opinion, the prevailing approaches, despite of using blockchain, are all centralized in nature, based on a full declaration of invoices put available online for their graph analysis. We differently advocate for a bottom‐up approach, a sort of agent approach, which the decentralized implementations are more appropriate, that fits nice and straightforward with Blockchain / Distributed Ledger Technologies (DLTs).

Distributed Intelligence algorithms, multi agent paradigms, require a proper framework that the computing and ledger possibilities of DLT platforms like Ethereum or Polygon, Polkadot, BitcoinSV, and Binance Smart Chain among others are ideal for developing such intelligence. Many distributed bottom‐up behaviours and strategies are open for research; coalition formation techniques and distributed autonomous organizations that require the proper platforms and environments to develop fully. There are actors like \cite{Invoicemate2022} that uses the BitcoinSV platform or \cite{finpay2022} that use blockchain to enhance the PoE.

\subsection{On Collateral}

Our solution requires no third parties and no collateral (but invoices are the basic asset) and is anti-cyclical, similarly to the complementary currencies like WIR Bank\footnote{\url{https://wwww.wir.ch}} \cite{de2015velocity}. In fact, currencies are the oldest information system of humanity, and today we have the means to create new information systems by new forms of payment that will complete the strength of the currencies.

However, using collateral and liquidity pools will indeed boost the invoice discounting, factoring or byppay adoption. Thus, in this project we’ll cover the areas of distributing intelligence with a DLT, and adopt new DeFi (Decentralized Finances) to cover risks or enhance liquidity of the invoices as assets, Real-world assets, tokenized on a ledger.

The DeFi implementation is backed by the ideas of value on the Internet of Value (IoV), where tokenized invoices back the value and their rights are traded among peers. With our new implementation of collateral, some of the IoV Systemic Risks are solved, remarkably the overcollateralization will be not necessary, as stated in our contributions to the recent report of \cite{delaRosaEsteva2022}.

\subsection{Liquidity Pools}

About Liquidity Pools for Crowd Funding of Factoring, there are Protocols for Loanable Funds (PLF) that let users borrow/lend digital assets in a decentralized fashion \cite{xu2022short}. Automated smart contracts behave as middle‐men. They lock assets deposited by the lender and allow borrowers to get liquidity in exchange for collateral. These types of smart contracts are also called Lending Pools [8]. These Lending Pools typically lock a pair of tokens, a loanable token, and a collateral token. By providing liquidity, lenders gain interest rates depending on the supply \& demand. Because there is no guarantee of paying back, Borrowers must over‐collateralize their position. On top of that, when returning the amount borrowed, the borrowers must pay an interest rate that is split pro‐rata among the lenders and the governance token. Moreover, when borrowers get liquidated, they will have to pay an additional fee.

\begin{figure}[H]
\includegraphics[width=.4\textwidth]{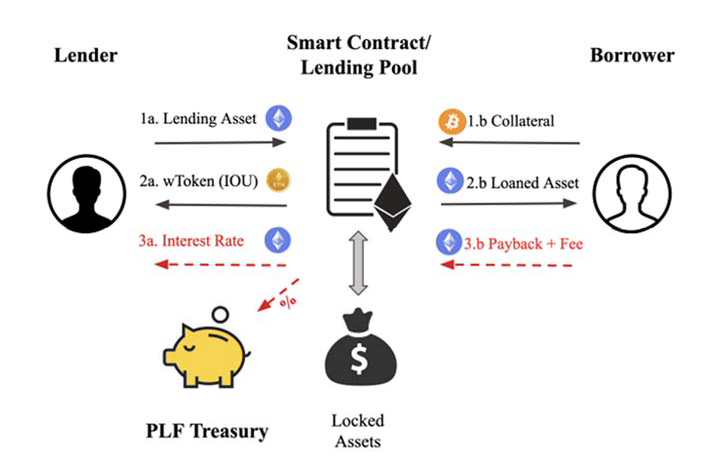}
\centering
\caption{\centering
Lending protocol framework \cite{xu2022short}
\label{fig:pools}}
\end{figure}  

Figure \ref{fig:pools} abstracts and generalizes the lending protocol framework by showing the main actors and interactions. From left to right, Lenders can deposit their crypto‐assets, Ethereum in this case, to gain additional profits. They receive a PLFs wrapped token or IOU as proof of their deposit. In the centre, the smart contract acts as a middle‐man. It takes care of the deposited assets, loans, and liquidations ‐ if any. On the right, a borrower must deposit collateral before getting the loan. Finally, at the end of the loan, the borrower will have to return the borrowed amount plus an interest rate, part of this interest rate will split pro‐rate among all lenders, and the rest will generate revenue for the PLF itself.

\begin{table}[H]
\caption{DeFi Protocols Comparison.\label{tab:liquidity_pools}}
  \centering
  \begin{tabular}{|c|c|c|c|c|}
			\toprule
			\textbf{DeFi Protocol} & \textbf{Smart Contract} & \textbf{Investor} & \textbf{User} & \textbf{Financial Service} \\
			\midrule
			\textit{Protocols for Loanable} & \multirow{2}{*}{Lending Pool} & \multirow{2}{*}{Lender} & \multirow{2}{*}{Borrower} & \multirow{2}{*}{Loan} \\
		      \textit{ Funds (PLF)} &  &  & & \\\midrule
            \textit{Decentralized Exchange} & \multirow{2}{*}{Liquidity Pool} & \multirow{2}{*}{Liquidity Provider} & \multirow{2}{*}{Buyer/Trader} & \multirow{2}{*}{Exchange} \\
            \textit{(DEX)} &  &  &  &  \\\midrule
            \textit{Yield Aggregators} & Vault & Vault User & - & Asset Management \\
			\bottomrule
  \end{tabular}
  \label{tab:table}
\end{table}

\subsection{From overcollateralization to risk appraisal}

The current initiatives on liquidity pools and related DeFi implementations, where whatever asset that is tokenized (i.e. lending, factoring, insurances, guaranties of service, creative industries, and more) is backed by collateral in the search of collateralizing the risk, by means of virtual currencies or any of the new digital assets \cite{delaRosaEsteva2022}. Their drawback lies on the overcollateralization, and their volatility caused by new forms of inter-dependency at an unprecedented level is a systemic risk in the coming internet of value.

In our opinion, the overcollateralization does not fit properly to invoices risk management, as they are assets with value already appraised in a currency and their risk is related to their liquidity. The euros that will be paid through an invoice is a sort of a stablecoin, invoice-euro for example, so that the conversion invoice-euro is pegged to the euro at a discount, say 0.5, 1, 2 percent or more according to the mora or invoice nonpayment rates. It is not a monetary risk, neither a crazy market valuation, but pure risk management, that is well known in financial markets which use sophisticated tools.

Differently from the solutions of liquidity pools in the state of the art that use crowdsource approaches to rate risk, we look for alternative formulas that appraise the risk in a fully automated way.

Under this reality, there is room to study partly collateralised invoices novel mechanisms, using risk management tools that are worth to explore and create in a decentralized implementation that should avoid overcollateralization. These techniques, like the Black-Scholes or the Kelly Criterion \cite{Kelly1956}, are being today studied for their implementation in risk appraisal and management. We are interested in the Kelly Criterion for calculating the optimal diversification of investments for creating a portfolio, given a volume of money available for investment, and given some known risk, as it forecasts the optimal benefit achievable.

In the case of decentralized markets with liquidity pools for invoices, provided that we know in advance the volume of money because it is the liquidity pool’s amount of money, we chose to reformulate the Kelly Criterion so that one might decide whether a given invoice that is partly collateralized can be accepted to get fully collateralized by the liquidity pool, and at what benefit to collect from to compensate the risk. This benefit will be the premium required as a fee for every invoice accepted in the liquidity pool. As this is a new usage of the Kelly Criterion, we named it Reverse Kelly Criterion as a new formula that calculates the premium to be granted given an invoice with some collateral, as we’ll see in the following section with all details.


\section{Reverse Kelly AMM (rkAMM) Design}

Electronic markets use algorithms that decide the price and amount of liquidity supplied in markets. But this automation is not complete as algorithms are managed by humans that intervene to improve their design, or turn them off entirely if conditions become unfavorable, leading to withdrawals of liquidity that may exacerbate market volatility.

Automated Market Maker (AMM) is a market, instead, consisted of just a single algorithm for facilitating trades. One that is simple, static and deterministic \cite{o2022can}. AMM follows a constant product rule for swapping assets: if the pool holds two types of assets then the product of the reserves of those assets must be the same before and after any swap is realized. This notion of keeping the value of a function (e.g., the product) of asset reserves invariant to swaps was later generalized into the idea of constant function market making. The basic construction of constant function market makers was formalized in \cite{angeris2020improved}. Then, a simple formula, $x \cdot y = k$, sets prices and quantities for thousands of different assets using only 378 lines of code. The AMM today \cite{o2022can} executes upwards of 50bn USD per month in digital assets. Only Uniswap, the largest of the AMMs, using a year of data acquired directly from the public Ethereum blockchain containing over 39 million transactions.

In an AMM, anyone can become a Liquidity Provider (LP) to the AMM by transferring their digital assets to pools. They might add to the liquidity of the pools, thereby increasing the pool size, or create new ones. In return they receive revenue from a fixed fee on trades (called swaps) by liquidity demanders to the AMM. Fee revenue is shared equally amongst LPs in proportion to their investment — so LP returns vary as a function of the amount of liquidity in a pool (or the pool’s volume or size). LPs can only add or remove liquidity, prices are set by the AMM algorithm, unlike traditional market makers that can also vary prices \cite{o2022can}.

An algorithm has been conceived following the main guidelines of an AMM for the invoices full collateralization. As said, AMM were designed to trade two currencies by their balance $x$ and $y$ in a liquidity pool to keep up a constant $k$, as $x \cdot y = k$ so that incoming $x$ means withdrawing $y$ to keep the $x \cdot y$ constant at $k$. There are several variants of the same scheme \cite{bichuch2022axioms} to add resilience and utility to the AMM implementation, that must follow their axioms. 

In our implementation we have as well liquidity providers that contribute and withdraw their funds at a fee, so that the liquidity pool has some volume $V$ to grant the remaining collateral demanded by the invoices to get fully funded, and premium is required in exchange to compensate the risk aiming at some benefit. From this point of view, it works as a standard community fund for securing risk of participants, but with liquidity providers instead and securing particular assets like invoices that are backed by some collateral. Thus collateral c1 goes in pool along with the premium $p$ and in exchange the remaining collateral c2 goes out, $V \cdot p = k (V)$; meaning that the constant $k$ is recalculated by the kelly criterion, thus $k$ changes with new liquidity volume $V$, and thus with the premium collected. 

As briefly introduced, we use the Kelly criterion to calculate the premium to be contributed to the liquidity pool after the collateral required for 100\% backing of an invoices that is already partly backed by some collateral. It is named rkAMM, after the name of Reverse Kelly AMM . Performance curves for the rkAMM have been analysed and simulated. In its implementation, we have considered the creation of NFTs with the inclusion of terms such as "backing \%" of the invoices, risk, elaboration, useful for applying the Kelly Criterion investment formula, and other considerations.

We assumed that for an invoice of amount $X$, the collateral $q$ is a percentage of $X$, and the remaining non‐collateralized amount is $p = 1 - q$.

The Kelly criterion (or Kelly strategy or Kelly bet), is a formula that determines the optimal theoretical size for a bet. It is valid when the expected returns are known. The Kelly bet size is found by maximizing the expected value of the logarithm of wealth, which is equivalent to maximizing the expected geometric growth rate. It was described by \cite{Kelly1956}. The criterion is also known as the scientific gambling method, as it leads to higher wealth compared to any other strategy in the long run (i.e. the theoretical maximum return as the number of bets goes to infinity). The practical use of the formula has been demonstrated for gambling and the same idea was used to explain diversification in investment management\footnote{\url{https://en.wikipedia.org/wiki/Kelly_criterion}}.

We use the Kelly criterion calculating how much premium must be required to contribute to a liquidity pool of size $V$ for an $X$ invoice amount with a collateral of $p$ thus it is demanded the remaining $q=1-p$ collateral out of the rkAMM.

\begin{figure}[H]
\includegraphics[scale=1]{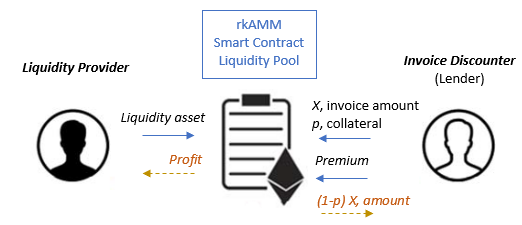}
\centering
\caption{\centering
The rkAMM input and output.
\label{fig:rkAMM_output}}
\end{figure}

Whenever the byppay is created, the Proof of Existence (PoE) of an invoice considers the mutual knowledge among A and C through B, the common acquaintance of A and C. This PoE and A rating must be enough for most of the Byppay cases that require no early payments.

If the Byppay is open to 3rd parties other than A, B, and C, then some collateral might be required to B to enhance the PoE of the AB invoice, and the collateralization flow might let the door open to third parties that trust B, or trust A, or trust both, to add collateral to that invoice, privately or openly as a Marketplace.

\begin{figure}[H]
\includegraphics[width=0.8\textwidth]{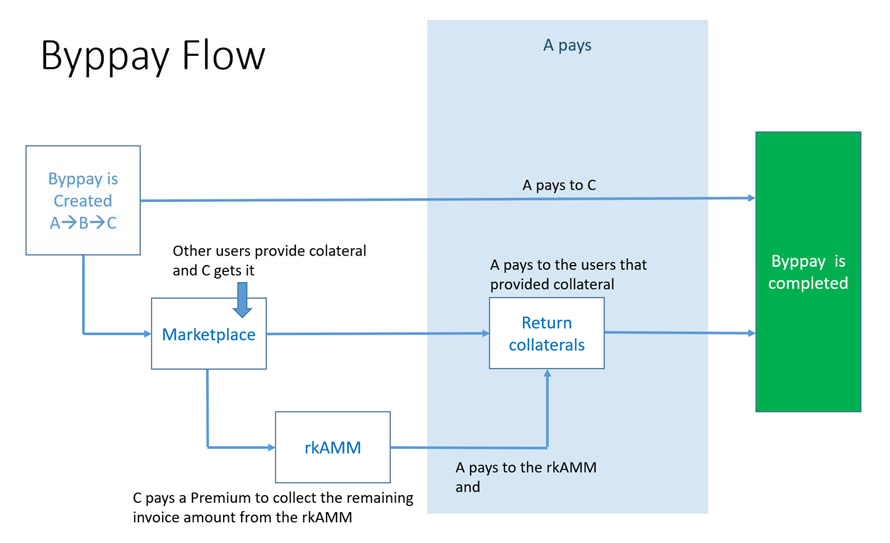}
\centering
\caption{\centering
How the rkAMM is included in the byppay platform.
\label{fig:byppay_flow}}
\end{figure}

Everyone contributing with collateral will be receiving a 2\% fee out of the piece of contribution to the collateral of the byppaying invoices. Invoices will be paid in 1 to 3 months on average, giving an average APY of 8\% to 24\%. The collateral contributors will know the due time of the invoices in advance to calculate their benefits. Also, the farthest the connection and knowledge on A and B, the more they would need oracles for rating/scoring of companies, which might fit well on this stage of the flow.

For those byppayable invoices with strong collateralization (say from 50\% collateral), they are eligible to the rkAMM, where C secures the remaining collateral out of a rkAMM liquidity pool out of a premium contributed to said liquidity pool so that the full amount of the invoice can be earned in advance of the payment due time of the AB invoice by A.

Whenever A pays, then all collateral to the Marketplace collateral providers is returned to the lenders at the 2\% extra, and the rkAMM liquidity pool gets back the collateral provided again at a 2\% extra, as well as B gets back the collateral he might have contributed at the 2\% extra, discountable to the fees and bonuses he might have released away for the whole Byppay process

\subsection{rkAMM Definition}
A model inspired by AMM is proposed to automatically fulfil the compensation of invoices. It is an underlying protocol to conduct exchanges in a decentralized manner through an autonomous trading mechanism. This allows users to exchange their assets without involving centralized financial authorities. We will use risk assessment criteria of Kelly formula.

In the proposed AMM, transactions will be executed considering the following two assets defined by:

\begin{itemize}
    \item The reserve for collateral allocated in the AMM as a liquidity pool, known as $Q$.
    \item The reserve of premium collected in the AMM, known as $Premium$.
\end{itemize}

The collateral serves as a liquidity asset used for completing the compensation of invoices to pay while in general terms the premium serves as a protection mechanism of the AMM against the loss of lent collateral and, hopefully, some benefit. Thus, the $premium$ is considered the amount to be paid based on the risk the AMM assumes when lending collateral.

These assets ($Q$ and $Premium$) are exchanged among agents by interacting with the AMM. Agents withdraw liquidity $Q$ at a premium $P$, to compensate the non‐collateralized amount of their invoices. This premium contributed by agents is used to compensate the losses the AMM may have during its life cycle. And the steady remaining premium will be considered a benefit attributed to the intrinsic behaviour of the AMM. As well, third parties, the liquidity providers, are allowed to deposit liquidity for $Q$ into the AMM which allow them to get a proportional profit out of the accumulated premium based on their shares on the AMM liquidity pool $Q$. Strategies to decide when and how much of the premium can be withdrawn are discussed later in following sections.

\begin{equation*}
    Profit_{Premium} = Collected_{Premium} - Collateral_{UnpaidInvoices}
\end{equation*}

Recall that the exchange of liquidity for premium happens instantaneously, while obtaining premium for those thirds who have deposited liquidity to the AMM will be an exchange that will be conducted eventually.

Also, it is important to clarify that if there is no liquidity available to collateralize an invoice, the AMM alternative will be to collateralize the invoice with the collected premium. This premium, if sufficient, will collateralize the invoice. The remaining premium as profit is expected to be shared in different percentages. However, in the proposed model it is suggested that there is a joint premium withdrawal simulating this action in a variable time period.

\subsection{rkAMM Formula}\label{rk_formula}

The formula on which the exchanges of collateral and premium are based will be here explained.
To do this, we start from Kelly's investment formula (Equation \ref{eq:1}) to work out the interpretation of its variables to adapt the formula to our decentralized invoice collateral service:

\begin{equation}\label{eq:1}
    f = \dfrac{p}{a} - \dfrac{q}{b}
\end{equation}

Where the following interpretations and assumptions are made:

\begin{itemize}
    \item $f$ is the ratio between the amount of collateral requested by the user and the total volume of the AMM. The total volume of the AMM is computed as the sum of available collateral Q and accumulated Premium at that time.
    \item $p$ is the collateralized percentage of an invoice amount. We considered that $p$ is the probability of the invoice to be paid.
    \item $q$ is the non‐collateralized percentage of an invoice amount. We considered that $q$ is the probability of the invoice to be unpaid.
    \item $a = q$ given that $q$ is the lack of collateral of an invoice to be fully collateralized by the AMM, we interpreted it as the amount of money to be lost in the operation, that is, $a$ which is known as the partial loss in Kelly’s investment formula.
    \item From the above: $p = 1 - q$
    \item $b$ in Kelly’s investment formula is defined as the partial win. We interpreted this value as the ratio between potential profit and the potential loss.
\end{itemize}

Simplifying the Equation \ref{eq:1} using the assumptions above:

\begin{equation}\label{eq:2}
    f = \frac{1 - q}{q} - \frac{q}{b}
\end{equation}

For our purpose, the variable to clear is $b$. Thus, solving $b$ from Equation \ref{eq:2}:

\begin{equation}\label{eq:3}
    b = \frac{q^{2}}{1 - q(f+1)}
\end{equation}

To clarify, the ratio $f$ and probabilities $p$ and $q$ are always known within an invoice. Ratio $f$ can be interpreted in diverse ways, but always keeping the same meaning:

\begin{equation}\label{eq:4}
    f = \frac{NonCollateralized_{Amount}}{{Volume_{AMM}}} = \frac{q  \cdot {Invoice_{TotalAmount}}}{{Volume_{AMM}}} = \frac{Q}{{Volume_{AMM}}}
\end{equation}

From now on, it is important to be aware that the value of $Q$ represents an amount calculated from the probability $p$ and the total amount of the invoice. Therefore, $Q$ should not be confused with $q$.

Also, $b$ which is calculated from Equation \ref{eq:3} is defined as the ratio between potential profit and the potential loss. The potential loss is known since it is the amount of collateral that the AMM lends to the user, that is, $Q$.

Finally, the variable to clear is the potential profit (i.e., $premium$) since it is the amount the AMM can earn in the operation if the user repays the collateral. That is why it is called potential profit, due to the AMM is assuming a risk and can lose the collateral if it is not repaid by the user.

\begin{equation}\label{eq:5}
    b = \frac{{PotentialProfit}}{{PotentialLoss}} = \frac{Premium}{Q}
\end{equation}

\begin{equation}\label{eq:6}
    Premium = b \cdot Q
\end{equation}

\subsection{Illustrative Example}

This section presents an illustrative example with calculations that could be common in a real scenario.

First, let’s declare the AMM initial conditions:

\begin{table}[H]
\caption{Initial rkAMM situation.\label{rkexample_1}}
\newcolumntype{C}{>{\centering\arraybackslash}X}
\begin{tabularx}{\textwidth}{|C|C|}
\toprule
\textbf{AMM initial condition} & \textbf{Value (euro)} \\
\midrule
        Liquidity available & 1,800 \\\midrule
        Premium deposited & 0 \\\midrule
        Volume (Liquidity + Premium) & 1,800 \\
\bottomrule
\end{tabularx}
\end{table}

Then, an invoice that has a $p$ and a $q$ over a total amount is applying for the AMM service:

\begin{table}[H]
\caption{Invoice amounts description.\label{rkexample_2}}
\newcolumntype{C}{>{\centering\arraybackslash}X}
\begin{tabularx}{\textwidth}{|C|C|C|}
\toprule
\textbf{Invoice} & \textbf{\%} & \textbf{Amount (euro)} \\
\midrule
        Collateralized amount ($p$) & 60 & 1,200 \\\midrule
        Non‐collateralized amount ($q$) & 40 & 800 \\\midrule
        Total & 100 & 1,200 \\
\bottomrule
\end{tabularx}
\end{table}

Using Equation \ref{eq:4}, the value of $f$ is:

\begin{equation*}
    f = \frac{Q}{Volume_{AMM}} = \frac{800}{1800} = 0.444...
\end{equation*}
    
Since $Q = q \cdot Invoice_{TotalAmount} = NonCollateralized_{Amount}$.

Then, from Equation \ref{eq:3}, the value of $b$ is:

\begin{equation*}
    b = \frac{q^{2}}{1 - q(f+1)} = \frac{0.4^{2}}{1 - 0.4(0.444+1)} = 0.3789 \rightarrow 37.89\%
\end{equation*}

Recall that $Q$ is proportional but essentially different from $q$ because with $p$, $q$, and $b$ and $a$, we are working with probabilities and ratios.

And from Equation \ref{eq:5} we have:

\begin{equation*}
    b = \frac{Premium}{Q} = \frac{Premium}{800} = 0.3789
\end{equation*}

Clearing premium from the previous equation as in Equation \ref{eq:6}:

\begin{equation*}
    Premium = 800 \cdot 0.3789 = 303.16 \text{ (euro)}
\end{equation*}

Finally, the AMM situation is the following:

\begin{table}[H] 
\caption{Intermediate rkAMM situation after accepting an invoice.\label{rkexample_3}}
\newcolumntype{C}{>{\centering\arraybackslash}X}
\begin{tabularx}{\textwidth}{|C|C|}
\toprule
 \textbf{AMM intermediate condition} & \textbf{Value (euro)} \\
\midrule
        Liquidity available & 1,000 \\\midrule
        Premium deposited & 303.16 \\\midrule
        Volume (Liquidity + Premium) & 1,303.16 \\
\bottomrule
\end{tabularx}
\end{table}

At this point, collateral is withdrawn from the AMM and $premium$ has been deposited. It is conceivable to wonder what will happen if the user does not repay any collateral, that will be matter of study with several scenarios that will be covered in the next section. Here the ideal and common functionality of the AMM is described.

So, following with this example, at some point the user or his client will repay the collateral, which is eight hundred euros. Finally, the AMM situation is:

\begin{table}[H] 
\caption{Final rkAMM situation after collateral repaid.\label{rkexample_4}}
\newcolumntype{C}{>{\centering\arraybackslash}X}
\begin{tabularx}{\textwidth}{|C|C|}
\toprule
\textbf{AMM final condition} & \textbf{Value (euro)} \\
\midrule
        Liquidity available & 1,800 \\\midrule
        Premium deposited & 303.16 \\\midrule
        Volume (Liquidity + Premium) & 2,103.16 \\
\bottomrule
\end{tabularx}
\end{table}

The AMM has made profit because it has assumed the risk by supporting the non‐collateralized amount of the invoice for some time in between the collateral is provided by the AMM and the collateral is later repaid within the invoice due time, sooner or later.

\section{Simulations Setup}

A set of simulations have been developed based on several scenarios to study typical situations that may occur. First, invoices used in scenarios have the following structure:

\begin{table}[H] 
\caption{Invoice attributes definition.\label{tab:invoice_def}}
\newcolumntype{C}{>{\raggedright\arraybackslash}X}
\begin{tabularx}{\textwidth}{|C|C|}
\toprule
\textbf{Value} & \textbf{Description} \\
\midrule
        Id Invoice & Invoice identifier \\\midrule
        Non‐coll. \% & Value between 5\% and 49\% \\\midrule
        Non‐coll. amount & Value between 100 and 2,000 (euro) \\\midrule
        Collateralized date & Day when the invoice is collateralized \\\midrule
        Collateralized status & True if the invoice is collateralized / False otherwise \\\midrule
        Payment delay days & Value between 30 and 120 (days) \\\midrule
        Paid status & True if the non‐coll. amount is repaid / False otherwise \\
\bottomrule
\end{tabularx}
\end{table}

Then, simulations results are based on a set of values. However, it must be noted that only few key attributes are modified for each scenario in order to interpret the behavior of the AMM in specific situations. The complete set of attributes used to simulate the scenarios is shown in Table \ref{tab:sim_param}.

\begin{table}[H]
\caption{Attributes used to simulate default and hack scenarios.\label{tab:sim_param}}
\newcolumntype{C}{>{\raggedright\arraybackslash}X}
\newcolumntype{P}[1]{>{\centering\arraybackslash}p{#1}}

			
   \begin{tabularx}{\textwidth}{|C|P{2.7cm}|C|P{2.8cm}|}

   \toprule
       \multirow{3}{*}{\textbf{Description}} & \multirow{3}{*}{\textbf{Value}} & \multirow{3}{*}{\textbf{Comment}} & {\textbf{Fixed or variable parameter (across different scenarios}}\\
			\midrule
			\multirow{3}{*}{Number of simulations} &	\multirow{3}{*}{100}	& Number of simulations to determine the common behaviour of the AMM & \multirow{3}{*}{Fix} \\\midrule
   
            \multirow{2}{*}{Initial collateral $Q_0$} & \multirow{2}{*}{10,000 (euro)} & This amount is the initial liquidity in the LP. & \multirow{2}{*}{Fix} \\\midrule
            
            Initial premium	& 0 & N/A & Fix \\\midrule
            
            Number of invoices & 500 & Max. value of invoices to process & Fix \\\midrule
            
            \multirow{3}{*}{\thead[l]{Range of \% from invoices to \\ collateralize}}	& \multirow{3}{*}{5\% - 49\%} & \multirow{3}{*}{\thead[l]{Max. and min. values allowed \\ for collateralization of invoices}} & Variable only for scenario 5 and Hack scenarios \\\midrule
            
            \multirow{6}{*}{\thead[l]{Range of amount from invoices \\ to collateralize }} & \multirow{6}{*}{100 – 2,000 (euro)} & Max. and min. amounts allowed to be collateralized. As the Initial collateral $Q_0$ is fixed to 10,000, we are instructing the simulations to be in the range 0,1\% - 20\% of the initial liquidity & \multirow{6}{*}{\thead[c]{Variable only for \\ scenario 4}} \\\midrule
            
            \multirow{2}{*}{Invoice payment period range} & \multirow{2}{*}{30 - 120 (days)} & Max. and min. delays in days for the payment of invoices & Variable only for scenario 3 \\\midrule
            
            \multirow{3}{*}{\thead[l]{Range of liquidity contribution \\ by LPs}} & \multirow{3}{*}{0 - 0} & Max. and min. amounts of liquidity contribution to the AMM & \multirow{3}{*}{\thead[c]{Variable only for \\ scenario 1}} \\\midrule
            
            \multirow{3}{*}{Non-payment probability} & \multirow{3}{*}{0\%} & The probability that collateral on an invoice will never be paid back & Variable only for scenario 2 and Hack scenarios \\\midrule
            
            \multirow{4}{*}{Hack probability} & \multirow{4}{*}{0\%} & The probability of including false invoices that will never be paid may affect the performance of the AMM &  \multirow{4}{*}{\thead[c]{Variable only for \\ Hack scenarios}} \\\midrule
            
            Probability of contribution by LPs & 0\% & The probability that an LP makes a liquidity deposit to the AMM  & Variable only for scenario 1 \\\midrule
            
            \multirow{2}{*}{Maximum days} & \multirow{2}{*}{500} & Days on which invoice entries are allowed & Fix to the number of invoices \\\midrule
            
            \multirow{3}{*}{Additional days} & \multirow{3}{*}{30} & Further days for simulations to observe the return to the steady state of the AMM & \multirow{3}{*}{Fix} \\\midrule
            
            \multirow{3}{*}{Number of days of the simulation} & \multirow{3}{*}{650} & Total days of simulation. Max. days + Max. delay invoice payment + Additional days & Fix to maximum (days + delays) + additional days \\
			\bottomrule
		\end{tabularx}
\end{table}

Please note that the simulations are based on uniformly random values and running a new simulation could give slightly different results. That is why there is a number of simulation parameter which we consider enough to show the behavior of the AMM.

\newpage

\subsection{Scenarios}\label{scenarios}

Initially and under normal circumstances, the following scenarios are considered to conduct the simulations:

\underline{Scenario 1} – There is a contribution of collateral by liquidity providers (LP) based on a growing rate concerning the volume of collateral ($Q$) in the AMM. The attributes modified in this scenario are not only the Probability of contribution fixed at 50\%, but also the Range of liquidity contribution by LPs in three different conditions: 1\% of initial $Q$ for scenario 1.1 (Very Low), 5\% of initial $Q$ for scenario 1.2 (Low), 10\% of initial $Q$ for scenario 1.3 (Medium) and 25\% of initial $Q$ for scenario 1.4 (Max).

\underline{Scenario 2} – Non-collateralized amount of a set of invoices is not repaid to the AMM. The only attribute changed in this scenario is the non-payment probability, defining three possible situations: 2\% of invoices for scenario 2.1 (Low), 5\% for scenario 2.2 (Medium), 20\% for scenario 2.3 (Max).

\underline{Scenario 3} - Increasing delay in invoice payments. Modify the payment period range of invoices from what is considered short to long. Payment period range: 30-60 days for scenario 3.1 (Short), 60-90 days for scenario 3.2 (Medium), 90-120 days for scenario 3.3 (Long).

\underline{Scenario 4} – Amount to be collateralized depends on a variable percentage of the initial volume of collateral ($Q_0$) in the AMM. Only the range of demanded collateral is modified in this scenario. Range of demanded collateral, thus liquidity contribution granted by LPs: 1\% of initial $Q_0$ for scenario 4.1 (Low), 10\% of initial $Q_0$ for scenario 4.2 (Medium), 25\% of initial $Q_0$ for scenario 4.3 (Max).

\underline{Scenario 5} – Different \% to be collateralized  with the same amount of invoices. Modify the range of \% collateral $p$ from invoices through three possible cases. Range of \% collateral $p$:  55\% for scenario 5.1 (Low), 75\% for scenario 5.2 (Medium), 90\% for scenario 5.3 (Max).

In addition to the scenarios proposed above, there is an additional set consisting of those situations in which there may be malicious actors trying to take advantage of the AMM operation. These scenarios foresee a hack of the AMM, that is, there is a high number of false, bogus invoices. A false invoice is considered one for which the collateral given away by the AMM will never be paid back. The purpose of this hack from the attacker's point of view is to drain the AMM liquidity.

\underline{Hacking scenario} - To hack the rkAMM, the following attributes from the default scenario are modified to produce a set of different hacking scenarios: 

\begin{itemize}
    \item Range of \% from invoices to collateralize: 49\% (Max), 30\% (Medium) and 10\% (Low) for all possible hack probabilities.
    \item Non‐payment probability: 10\% (Low), 50\% (Medium) and 100\% (Max) for all previous ranges of \% from invoices to collateralize.
\end{itemize}




\subsection{Technical Description of the Simulator}

The simulator in charge of performing the simulations has been developed in Python. Additionally, for the analysis and presentation of the computed metrics common analysis and data visualisation libraries have been used.

The invoices have been structured in a list of dictionaries to execute the simulations. Each invoice is defined by a dictionary of seven keys and values, and this dictionary in turn occupies a position in a list of invoices.

Regarding the functions of the simulation, those that allow modifying the volumes of liquidity, collateral and premium have been defined. In addition, a function has been developed to calculate the premium based on the requested collateral, as explained in section \ref{rk_formula} of this document.

To implement the simulation there is a main function that randomly generates a set of invoices based on the previously defined parameters in Table \ref{tab:invoice_def}. The flow of invoices is steady at 1 invoice per day with uniformly random amount and collateral within the range of parameters described for all the simulations in Table \ref{tab:sim_param}. When they are accepted in the liquidity pool, then premium is contributed, and collateral is withdrawn at the same day. Otherwise, when they are not accepted, they are simply discarded, and the next day comes for a next invoice.

The simulation also synchronizes once a day the liquidity contributions of uniformly random amounts in the range established by the parameters. The invoices that are repaid they do it with a uniformly random day within the range set of parameters from a minimum number of days (often 30 days are chosen) and a maximum of 120 days. On the other hand, invoices that are not repaid they do it with a value that guarantees to be higher than the maximum delay, that is 100,000, since Python 3's \texttt{int} does not have a maximum size and using \texttt{math.inf} may impact the efficiency of our code. In this way we ensure that these invoices will not be considered in any scenario.

Then, the number of days for each simulation is calculated from the number of days for the flow of all invoices plus the maximum delay and an additional number of extra days. For instance, for 500 invoices, a maximum delay of 120 days, and 30 extra days, the limit is $500+120+30 = 650$ days per simulation.

Every scenario is defined by the mentioned parameters in Table \ref{tab:sim_param}, and a batch of 100 simulations per scenario are run, thus the metrics of the batch are calculated as an average of the index at all simulations.

To simulate several scenarios of unpaid invoices, a uniform random number of invoices will be chosen not to be repaid, so that they had paid the premium to be accepted in the AMM, but got the collateral that is never returned to the AMM. The AMM will try to compensate the losses by means of the premium and we’ll see how resilient is at the end of the run of all simulations.

Also, the AMM contemplates the possibility of withdrawing the premium collted from the users through a function. This mechanism can be adapted, both the withdrawal period and the premium percentage to be withdrawn can be modified.

And last, but not least, the hack of the system, consisting in extenuating the AMM by a flow of bogus invoices, intentionally not to be paid. This is done by increasing the hack probability, which in turn increases the non-payment probability of invoices.

\subsection{Repository of the rkAMM}
There is a GitHub repository\footnote{\url{https://github.com/ballesterosbr/rkAMM}} of the rkAMM for running and testing the simulations describe above.

\section{Results}

A batch of simulations that proved the behavior of the AMM based on the scenarios defined in section \ref{scenarios} of this document were successfully performed. For each scenario, those values that suffer the greatest variations according to the input values will be presented and discussed. Also, a set of curves will be presented for the most revelant scenarios.


\subsection{Limitations}

In the event that the AMM does not have liquidity nor premium to collateralize an invoice, this invoice is discarded and a next one will be evaluated. When the AMM gets a payback of the collateral granted to a former collateralized invoice or the amount to be collateralized of a new invoice is less than the available liquidity in the AMM, then the invoice might be granted for collateralization.

In the simulations, the profit percentage shows the result of the AMM. This benefit will be distributed among those LPs that have deposited liquidity into the AMM, so it is expected that this will have an impact on the AMM profit.

After inspecting the scenarios, our proposal of the AMM implies that the invoice percentage to collateralize is advised to be never more than 49\% or lower than 5\%.

\subsection{Metrics}

A set of metrics is defined to describe the behaviour of the AMM to analyse the results of the simulations, out of which, several curves will be plotted for the volume of the AMM, the premium collected, the premium withdrawn, and the resulting liquidity as an addition of the volume and the remaining premium. To better understand the parameters considered in the metrics, Table \ref{baseline_output} shows an output example. At the end of the series of 100 simulations, we have these averaged measures.

The set of Tables \ref{tab:rkAMM_default_results_1day}, \ref{tab:rkAMM_hack_results_1day}, \ref{tab:rkAMM_default_results_30days}, \ref{tab:rkAMM_hack_results_30days}, \ref{tab:rkAMM_default_results_90days}, and \ref{tab:rkAMM_hack_results_90days} show all scenarios results, which will be discussed in next section. The result curves are also presented for a better understanding of the results. To avoid overloading this section and to facilitate reading, only the curves from scenario 5 with a withdrawal period of 30 days are presented, since it is considered one of the most expected scenarios to occur in a real environment. The rest of the curves for the rest of scenarios can be found in the Appendix \ref{appendix} section.

\begin{table}[H] 
\caption{Baseline scenario output. Scenario 5 and 30 days of withdrawal period.\label{baseline_output}}

\newcolumntype{P}[1]{>{\centering\arraybackslash}p{#1}}
\newcolumntype{J}[1]{>{\raggedright\arraybackslash}p{#1}}
\newcolumntype{C}{>{\raggedright\arraybackslash}X}

\begin{tabularx}{\textwidth}{|J{4.1cm}|P{1.8cm}|P{1.8cm}|P{1.9cm}|C|}

\toprule
\multirow{3}{*}{\textbf{Value}} & {\textbf{No Premium Withdrawn}} & {\textbf{Premium Withdrawn}} & {\textbf{Premium vs. No Premium Withdrawn}} & \multirow{3}{*}{\textbf{Description}}\\
\toprule

\multirow{2}{*}{Number of simulations} & \multirow{2}{*}{100.0} & \multirow{2}{*}{100.0} & \multirow{2}{*}{0.0\%} & Number of simulations conducted. \\\midrule

\multirow{3}{*}{Simulation time period (days)} & \multirow{3}{*}{650.0} & \multirow{3}{*}{650.0} & \multirow{3}{*}{0.0\%} & Total number of days on which the simulation has been conducted. \\\midrule

\multirow{2}{*}{Total of invoices} & \multirow{2}{*}{500.0} & \multirow{2}{*}{500.0} & \multirow{2}{*}{0.0\%} & Number of invoices to be processed by the AMM. \\\midrule

Average of accepted invoices (collateralized) & \multirow{2}{*}{350.96} & \multirow{2}{*}{179.34} & \multirow{2}{*}{-48.9\%} & Number of invoices accepted by the AMM. \\\midrule

\% of accepted invoices (collateralized) & \multirow{2}{*}{70.19} & \multirow{2}{*}{35.87} & \multirow{2}{*}{-48.9\%} & \% of invoices accepted by the AMM. \\\midrule

Average of paid invoices (capital returned) & \multirow{3}{*}{350.96} & \multirow{3}{*}{179.34} & \multirow{3}{*}{-48.9\%} & Number of accepted invoices for which collateral has been returned. \\\midrule

\% of paid invoices (capital returned) & \multirow{2}{*}{100.0} & \multirow{2}{*}{100.0} & \multirow{2}{*}{0.0\%} & \% of accepted invoices for which collateral has been returned. \\\midrule

\multirow{3}{*}{\thead[l]{Average of unpaid invoices \\ (capital not returned) }} & \multirow{3}{*}{0.0} & \multirow{3}{*}{0.0} & \multirow{3}{*}{0.0\%} & Number of accepted invoices for which the collateral has not been returned. \\\midrule

\multirow{3}{*}{\thead[l]{\% of unpaid invoices (capital \\ not returned)}} & \multirow{3}{*}{0.0} & \multirow{3}{*}{0.0} & \multirow{3}{*}{0.0\%} & \% of accepted invoices for which the collateral has not been returned. \\\midrule

Average loss due to unpaid invoices) & \multirow{2}{*}{0.0} & \multirow{2}{*}{0.0} & \multirow{2}{*}{0.0\%} & Amount of Euro lost due to unpaid invoices. \\
\midrule

\multirow{2}{*}{Total collateral covered} & \multirow{2}{*}{329,583.29} & \multirow{2}{*}{124,772.17} & \multirow{2}{*}{-62.14\%} & Total amount of collateral borrowed by the AMM. \\\midrule

\multirow{3}{*}{\thead[l]{Total collateral covered \\ x (i.c.))}} & \multirow{3}{*}{32.96} & \multirow{3}{*}{12.48} & \multirow{3}{*}{-62.14\%} & Total amount of collateral borrowed by the AMM with respect to the initial collateral $Q_0$. \\\midrule

Total premium withdrawn & 0.0 & 33,161.86 & 100.0\% & Total premium withdrawn. \\\midrule

Total premium withdrawn x (i.c.) & \multirow{2}{*}{0.0} & \multirow{2}{*}{3.32} & \multirow{2}{*}{100.0\%} & Total premium withdrawn with respect to the initial collateral $Q_0$. \\\midrule

\multirow{2}{*}{Remaining AMM premium} & \multirow{2}{*}{10,344.89} & \multirow{2}{*}{0.0} & \multirow{2}{*}{-100.0\%} & Remaining premium in the AMM after simulation. \\\midrule

\multirow{3}{*}{\thead[l]{Remaining AMM premium \\ x (i.c.))}} & \multirow{3}{*}{1.03} & \multirow{3}{*}{0.0} & \multirow{3}{*}{-100.0\%} & Remaining premium in the AMM after simulation with respect to the initial collateral $Q_0$. \\\midrule

\multirow{2}{*}{Final AMM volume} & \multirow{2}{*}{98,416.65} & \multirow{2}{*}{29,049.41} & \multirow{2}{*}{-70.48\%} & Total volume of the AMM after simulation. \\\midrule

\multirow{2}{*}{AMM profit} & \multirow{2}{*}{88,416.65} & \multirow{2}{*}{52,211.27} & \multirow{2}{*}{-40.95\%} & AMM profit without including initial collateral $Q_0$. \\\midrule

\multirow{2}{*}{AMM profit percentage} & \multirow{2}{*}{884.17} & \multirow{2}{*}{522.11} & \multirow{2}{*}{-40.95\%} & AMM profit percentage without including initial collateral $Q_0$. \\

\bottomrule
\end{tabularx}
\end{table}

\newpage

\newgeometry{left=2.5cm,top=0.4cm,bottom=0.3cm}

\begin{landscape}

\newcolumntype{P}[1]{>{\centering\arraybackslash}p{#1}}
\begin{table}[H]

\caption{rkAMM results. Every 1 day a 50\% of the premium obtained is withdrawn. \label{tab:rkAMM_default_results_1day}}
\begin{tabularx}{\linewidth }
{|P{1.3cm}|P{1.3cm}|P{1.3cm}|P{1.3cm}|P{1.3cm}|P{1.6cm}|P{1.6cm}|P{1.3cm}|P{1.3cm}|P{1.5cm}|P{1.3cm}|P{1.3cm}|P{1.3cm}|P{1.35cm}|P{1.35cm}|}

        \toprule
        Scenario & \multicolumn{2}{c|}{\% accepted invoices}& \multicolumn{2}{c|}{\% unpaid invoices}&  \multicolumn{2}{c|}{Avg. loss due to} & \multicolumn{2}{c|}{Total collateral covered} & \multicolumn{2}{c|}{Total premium with.}& \multicolumn{2}{c|}{Remaining premium} & \multicolumn{2}{c|}{AMM profit} \\
            
        \# & \multicolumn{2}{c|}{(collateralized)} & \multicolumn{2}{c|}{(capital not returned)}&             \multicolumn{2}{c|}{unpaid invoices} & \multicolumn{2}{c|}{x (i.c.)}&             \multicolumn{2}{c|}{x (i.c.)}& \multicolumn{2}{c|}{x (i.c.)}& \multicolumn{2}{c|}{percentage} \\
        \toprule

        & No  With. &  With. & No With. & With. & No With. & With. & No With. & With. & No With. & With. & No With. & With. & No With. & With. \\
        \toprule
        
        1.1 & 69.39 & 36.29 & 0.0 & 0.0 & 0.0 & 0.0 & 32.45 & 12.62 & 0.0 & 3.43 & 1.01 & 0.0 & 872.63 & 537.8 \\
        1.2 & 87.42 & 69.22 & 0.0 & 0.0 & 0.0 & 0.0 & 43.69 & 31.48 & 0.0 & 7.03 & 4.02 & 0.0 & 1,601.63 & 1,556.73 \\
        1.3 & 93.55 & 86.47 & 0.0 & 0.0 & 0.0 & 0.0 & 47.95 & 43.29 & 0.0 & 7.91 & 5.46 & 0.0 & 2,407.74 & 2,437.59 \\
        1.4 & 99.18 & 97.5 & 0.0 & 0.0 & 0.0 & 0.0 & 52.04 & 50.75 & 0.0 & 7.9 & 6.86 & 0.0 & 4,901.76 & 4,897.91 \\
        2.1 & 56.1 & 22.94 & 1.9 & 2.02 & 4,512.98 & 1,472.62 & 24.46 & 7.0 & 0.0 & 1.96 & 0.28 & 0.0 & 554.43 & 199.11 \\
        2.2 & 51.57 & 21.56 & 5.11 & 4.99 & 11,049.93 & 3,097.52 & 21.69 & 6.37 & 0.0 & 1.76 & 0.19 & 0.0 & 447.58 & 160.02 \\
        2.3 & 36.52 & 15.0 & 19.98 & 19.76 & 26,130.04 & 8,128.02 & 13.09 & 4.07 & 0.0 & 1.07 & 0.05 & 0.0 & 117.39 & 34.6 \\
        3.1 & 80.55 & 35.11 & 0.0 & 0.0 & 0.0 & 0.0 & 39.97 & 12.42 & 0.0 & 3.3 & 2.81 & 0.0 & 748.97 & 359.58 \\
        3.2 & 52.98 & 22.82 & 0.0 & 0.0 & 0.0 & 0.0 & 23.54 & 7.36 & 0.0 & 1.97 & 0.29 & 0.0 & 561.77 & 213.34 \\
        3.3 & 32.34 & 16.04 & 0.0 & 0.0 & 0.0 & 0.0 & 13.08 & 5.27 & 0.0 & 1.4 & 0.02 & 0.0 & 340.62 & 150.89 \\
        4.1 & 100.0 & 100.0 & 0.0 & 0.0 & 0.0 & 0.0 & 5.0 & 5.0 & 0.0 & 0.74 & 0.73 & 0.0 & 72.62 & 73.62 \\
        4.2 & 48.68 & 18.1 & 0.0 & 0.0 & 0.0 & 0.0 & 24.34 & 9.05 & 0.0 & 4.01 & 0.31 & 0.0 & 578.46 & 457.26 \\
        4.3 & 21.6 & 8.21 & 0.0 & 0.0 & 0.0 & 0.0 & 27.0 & 10.27 & 0.0 & 5.14 & 0.22 & 0.0 & 728.34 & 608.29 \\
        5.1 & 96.87 & 30.29 & 0.0 & 0.0 & 0.0 & 0.0 & 50.2 & 10.16 & 0.0 & 8.05 & 13.61 & 0.0 & 2,182.86 & 903.32 \\
        5.2 & 32.56 & 22.87 & 0.0 & 0.0 & 0.0 & 0.0 & 10.79 & 6.84 & 0.0 & 0.72 & 0.03 & 0.0 & 114.05 & 72.34 \\
        5.3 & 23.46 & 22.94 & 0.0 & 0.0 & 0.0 & 0.0 & 7.12 & 6.85 & 0.0 & 0.08 & 0.01 & 0.0 & 8.49 & 8.18 \\

\bottomrule
\multicolumn{5}{|c|}{\textit{*i.c. = initial collateral $Q_0$}} & \multicolumn{5}{|c|}{\textit{No With. = No Premium Withdrawn}} & \multicolumn{5}{|c|}{\textit{With. = Premium Withdrawn}} \\ 
\bottomrule
\end{tabularx}
\end{table}

\begin{table}[H]
\caption{rkAMM hack scenario results. Every 1 day a 50\% of the premium obtained is withdrawn.\label{tab:rkAMM_hack_results_1day}}
\begin{tabularx}{\linewidth }
{|P{1.3cm}|P{1.3cm}|P{1.3cm}|P{1.3cm}|P{1.3cm}|P{1.6cm}|P{1.6cm}|P{1.3cm}|P{1.3cm}|P{1.5cm}|P{1.3cm}|P{1.3cm}|P{1.3cm}|P{1.35cm}|P{1.35cm}|}

        \toprule
        Hack & \multicolumn{2}{c|}{\% accepted invoices}& \multicolumn{2}{c|}{\% unpaid invoices}&  \multicolumn{2}{c|}{Avg. loss due to} & \multicolumn{2}{c|}{Total collateral covered} & \multicolumn{2}{c|}{Total premium with.}& \multicolumn{2}{c|}{Remaining premium} & \multicolumn{2}{c|}{AMM profit} \\
            
        Prob. & \multicolumn{2}{c|}{(collateralized} & \multicolumn{2}{c|}{(collateralized)}&             \multicolumn{2}{c|}{unpaid invoices} & \multicolumn{2}{c|}{x (i.c.)}&             \multicolumn{2}{c|}{x (i.c.)}& \multicolumn{2}{c|}{x (i.c.)}& \multicolumn{2}{c|}{percentage} \\
        \toprule

        & No  With. &  With. & No With. & With. & No With. & With. & No With. & With. & No With. & With. & No With. & With. & No With. & With. \\
        \toprule

{\multirow{3}{*}{10 \%}} &  {99.94} &  {47.72} &  {9.78} &  {9.91} &  {51,136.77} &  {19,804.32} &  {52.3} &  {20.08} &  {0.0} &  {26.79} &  {16.33} &  {0.0} &  {2,300.96} &  {3,008.99 } \\
&  {32.63} &  {18.19} &  {10.16} &  {9.93} &  {10,627.94} &  {5,147.54} &  {10.52} &  {5.12} &  {0.0} &  {0.89} &  {0.04} &  {0.0} &  {80.26} &  {38.79 } \\
& \textcolor{purple} {19.16} & \textcolor{purple} {18.51} & \textcolor{purple} {9.8} & \textcolor{purple} {9.54} & \textcolor{purple} {5,211.17} & \textcolor{purple} {4,895.74} & \textcolor{purple} {5.38} & \textcolor{purple} {5.21} & \textcolor{purple} {0.0} & \textcolor{purple} {0.06} & \textcolor{purple} {0.01} & \textcolor{purple} {0.0} & \textcolor{purple} {-45.71} & \textcolor{purple} {-42.75 } \\
 {\multirow{3}{*}{50 \%}} &  {99.95} &  {13.96} &  {50.19} &  {50.13} &  {264,936.55} &  {22,607.54} &  {52.8} &  {4.52} &  {0.0} &  {6.55} &  {1.74} &  {0.0} &  {483.79} &  {572.29 } \\
& \textcolor{purple} {11.4} & \textcolor{purple} {6.68} & \textcolor{purple} {49.93} & \textcolor{purple} {49.64} & \textcolor{purple} {14,482.67} & \textcolor{purple} {9,884.92} & \textcolor{purple} {2.92} & \textcolor{purple} {1.96} & \textcolor{purple} {0.0} & \textcolor{purple} {0.32} & \textcolor{purple} {0.01} & \textcolor{purple} {0.0} & \textcolor{purple} {-94.93} & \textcolor{purple} {-66.01 } \\
& \textcolor{purple} {7.39} & \textcolor{purple} {7.11} & \textcolor{purple} {49.24} & \textcolor{purple} {48.86} & \textcolor{purple} {10,009.92} & \textcolor{purple} {9,823.79} & \textcolor{purple} {2.01} & \textcolor{purple} {2.0} & \textcolor{purple} {0.0} & \textcolor{purple} {0.02} & \textcolor{purple} {0.0} & \textcolor{purple} {0.0} & \textcolor{purple} {-97.73} & \textcolor{purple} {-95.88 } \\
\textcolor{purple} {\multirow{3}{*}{100 \%}} & \textcolor{purple} {98.99} & \textcolor{purple} {2.35} & \textcolor{purple} {100.0} & \textcolor{purple} {100.0} & \textcolor{purple} {515,511.86} & \textcolor{purple} {10,923.77} & \textcolor{purple} {51.55} & \textcolor{purple} {1.09} & \textcolor{purple} {0.0} & \textcolor{purple} {0.84} & \textcolor{purple} {0.33} & \textcolor{purple} {0.0} & \textcolor{purple} {-66.94} & \textcolor{purple} {-16.02 } \\
& \textcolor{purple} {2.75} & \textcolor{purple} {2.15} & \textcolor{purple} {100.0} & \textcolor{purple} {100.0} & \textcolor{purple} {11,719.05} & \textcolor{purple} {9,992.76} & \textcolor{purple} {1.17} & \textcolor{purple} {1.0} & \textcolor{purple} {0.0} & \textcolor{purple} {0.15} & \textcolor{purple} {0.01} & \textcolor{purple} {0.0} & \textcolor{purple} {-99.44} & \textcolor{purple} {-84.9 } \\
& \textcolor{purple} {2.27} & \textcolor{purple} {2.24} & \textcolor{purple} {100.0} & \textcolor{purple} {100.0} & \textcolor{purple} {10,077.26} & \textcolor{purple} {9,962.26} & \textcolor{purple} {1.01} & \textcolor{purple} {1.0} & \textcolor{purple} {0.0} & \textcolor{purple} {0.01} & \textcolor{purple} {0.0} & \textcolor{purple} {0.0} & \textcolor{purple} {-99.61} & \textcolor{purple} {-98.48 } \\

\bottomrule
\multicolumn{5}{|c|}{\textit{*i.c. = initial collateral $Q_0$}} & \multicolumn{5}{|c|}{\textit{No With. = No Premium Withdrawn}} & \multicolumn{5}{|c|}{\textit{With. = Premium Withdrawn}} \\ 
\bottomrule
\end{tabularx}
\end{table}
\vspace*{-5mm}


\newpage

\newcolumntype{P}[1]{>{\centering\arraybackslash}p{#1}}
\begin{table}[H]

\caption{rkAMM results. Every 30 days a 50\% of the premium obtained is withdrawn. \label{tab:rkAMM_default_results_30days}}
\begin{tabularx}{\linewidth }
{|P{1.3cm}|P{1.3cm}|P{1.3cm}|P{1.3cm}|P{1.3cm}|P{1.6cm}|P{1.6cm}|P{1.3cm}|P{1.3cm}|P{1.5cm}|P{1.3cm}|P{1.3cm}|P{1.3cm}|P{1.35cm}|P{1.35cm}|}

        \toprule
        Scenario & \multicolumn{2}{c|}{\% accepted invoices}& \multicolumn{2}{c|}{\% unpaid invoices}&  \multicolumn{2}{c|}{Avg. loss due to} & \multicolumn{2}{c|}{Total collateral covered} & \multicolumn{2}{c|}{Total premium with.}& \multicolumn{2}{c|}{Remaining premium} & \multicolumn{2}{c|}{AMM profit} \\
            
        \# & \multicolumn{2}{c|}{(collateralized)} & \multicolumn{2}{c|}{(capital not returned)}&             \multicolumn{2}{c|}{unpaid invoices} & \multicolumn{2}{c|}{x (i.c.)}&             \multicolumn{2}{c|}{x (i.c.)}& \multicolumn{2}{c|}{x (i.c.)}& \multicolumn{2}{c|}{percentage} \\
        \toprule

        & No  With. &  With. & No With. & With. & No With. & With. & No With. & With. & No With. & With. & No With. & With. & No With. & With. \\
        \toprule
        
1.1 & 69.64 & 66.23 & 0.0 & 0.0 & 0.0 & 0.0 & 32.5 & 30.24 & 0.0 & 1.5 & 0.96 & 0.01 & 868.36 & 860.4 \\ 
1.2 & 87.49 & 86.28 & 0.0 & 0.0 & 0.0 & 0.0 & 43.73 & 42.93 & 0.0 & 4.32 & 4.05 & 0.02 & 1,610.66 & 1,603.16 \\ 
1.3 & 93.97 & 93.16 & 0.0 & 0.0 & 0.0 & 0.0 & 48.18 & 47.8 & 0.0 & 5.78 & 5.48 & 0.02 & 2,432.65 & 2,428.88 \\ 
1.4 & 99.11 & 98.9 & 0.0 & 0.0 & 0.0 & 0.0 & 51.8 & 51.71 & 0.0 & 7.02 & 6.83 & 0.02 & 4,868.1 & 4,828.38 \\ 
2.1 & 56.42 & 50.38 & 2.16 & 2.06 & 5,418.73 & 4,547.89 & 24.66 & 21.2 & 0.0 & 0.7 & 0.25 & 0.0 & 561.84 & 501.42 \\ 
2.2 & 51.99 & 48.28 & 5.15 & 4.88 & 11,524.67 & 9,398.72 & 21.9 & 19.79 & 0.0 & 0.64 & 0.18 & 0.0 & 453.12 & 424.97 \\ 
2.3 & 37.41 & 32.0 & 19.79 & 19.9 & 27,238.25 & 21,592.14 & 13.65 & 10.88 & 0.0 & 0.36 & 0.06 & 0.0 & 131.24 & 97.31 \\ 
3.1 & 80.74 & 78.38 & 0.0 & 0.0 & 0.0 & 0.0 & 40.1 & 38.35 & 0.0 & 3.04 & 2.89 & 0.02 & 757.77 & 759.62 \\ 
3.2 & 52.49 & 49.01 & 0.0 & 0.0 & 0.0 & 0.0 & 23.35 & 21.08 & 0.0 & 0.75 & 0.31 & 0.01 & 561.81 & 529.75 \\ 
3.3 & 32.82 & 29.5 & 0.0 & 0.0 & 0.0 & 0.0 & 13.22 & 11.76 & 0.0 & 0.35 & 0.03 & 0.0 & 344.3 & 312.77 \\ 
4.1 & 100.0 & 100.0 & 0.0 & 0.0 & 0.0 & 0.0 & 5.0 & 5.0 & 0.0 & 0.74 & 0.73 & 0.0 & 72.53 & 73.77 \\ 
4.2 & 48.95 & 41.87 & 0.0 & 0.0 & 0.0 & 0.0 & 24.48 & 20.93 & 0.0 & 0.81 & 0.3 & 0.0 & 581.6 & 525.56 \\ 
4.3 & 21.88 & 15.94 & 0.0 & 0.0 & 0.0 & 0.0 & 27.34 & 19.93 & 0.0 & 1.39 & 0.21 & 0.01 & 728.41 & 584.32 \\ 
5.1 & 96.83 & 95.82 & 0.0 & 0.0 & 0.0 & 0.0 & 49.92 & 49.22 & 0.0 & 13.94 & 13.49 & 0.06 & 2,172.02 & 2,216.21 \\ 
5.2 & 32.11 & 30.23 & 0.0 & 0.0 & 0.0 & 0.0 & 10.77 & 9.81 & 0.0 & 0.23 & 0.03 & 0.0 & 113.91 & 103.73 \\
5.3 & 23.7 & 23.16 & 0.0 & 0.0 & 0.0 & 0.0 & 7.14 & 6.92 & 0.0 & 0.05 & 0.01 & 0.0 & 8.52 & 8.26 \\

\bottomrule
\multicolumn{5}{|c|}{\textit{*i.c. = initial collateral $Q_0$}} & \multicolumn{5}{|c|}{\textit{No With. = No Premium Withdrawn}} & \multicolumn{5}{|c|}{\textit{With. = Premium Withdrawn}} \\ 
\bottomrule
\end{tabularx}
\end{table}

\begin{table}[H]
\caption{rkAMM hack scenario results. Every 30 days a 50\% of the premium obtained is withdrawn.\label{tab:rkAMM_hack_results_30days}}
\begin{tabularx}{\linewidth }
{|P{1.3cm}|P{1.3cm}|P{1.3cm}|P{1.3cm}|P{1.3cm}|P{1.6cm}|P{1.6cm}|P{1.3cm}|P{1.3cm}|P{1.5cm}|P{1.3cm}|P{1.3cm}|P{1.3cm}|P{1.35cm}|P{1.35cm}|}

        \toprule
        Hack & \multicolumn{2}{c|}{\% accepted invoices}& \multicolumn{2}{c|}{\% unpaid invoices}&  \multicolumn{2}{c|}{Avg. loss due to} & \multicolumn{2}{c|}{Total collateral covered} & \multicolumn{2}{c|}{Total premium with.}& \multicolumn{2}{c|}{Remaining premium} & \multicolumn{2}{c|}{AMM profit} \\
            
        Prob. & \multicolumn{2}{c|}{(collateralized} & \multicolumn{2}{c|}{(collateralized)}&             \multicolumn{2}{c|}{unpaid invoices} & \multicolumn{2}{c|}{x (i.c.)}&             \multicolumn{2}{c|}{x (i.c.)}& \multicolumn{2}{c|}{x (i.c.)}& \multicolumn{2}{c|}{percentage} \\
        \toprule

        & No  With. &  With. & No With. & With. & No With. & With. & No With. & With. & No With. & With. & No With. & With. & No With. & With. \\
        \toprule

{\multirow{3}{*}{10 \%}} &  {99.96} & {99.64} & {9.86} & {9.77} & {51,256.79} & {50,736.8} & {52.45} & {52.32} & {0.0} & {17.67} & {16.46} & {0.07} & {2,309.78} & {2,442.9 } \\

& {32.12} & {29.94} & {10.14} & {9.83} & {10,675.98} & {9,172.03} & {10.39} & {9.48} & {0.0} & {0.29} & {0.04} & {0.0} & {77.82} & {76.84 } \\

& \textcolor{purple} {18.65} & \textcolor{purple} {18.08} & \textcolor{purple} {10.07} & \textcolor{purple} {10.52} & \textcolor{purple} {5,323.51} & \textcolor{purple} {5,332.92} & \textcolor{purple} {5.2} & \textcolor{purple} {5.05} & \textcolor{purple} {0.0} & \textcolor{purple} {0.04} & \textcolor{purple} {0.01} & \textcolor{purple} {0.0} & \textcolor{purple} {-47.04} & \textcolor{purple} {-47.31 } \\

{\multirow{3}{*}{50 \%}} &  {99.95} & {99.43} & {50.21} & {50.05} & {264,606.12} & {261,539.02} & {52.69} & {52.11} & {0.0} & {6.09} & {1.84} & {0.02} & {483.06} & {904.98 } \\

& \textcolor{purple} {11.31} & \textcolor{purple} {9.8} & \textcolor{purple} {49.37} & \textcolor{purple} {50.69} & \textcolor{purple} {14,483.85} & \textcolor{purple} {13,215.62} & \textcolor{purple} {2.93} & \textcolor{purple} {2.66} & \textcolor{purple} {0.0} & \textcolor{purple} {0.11} & \textcolor{purple} {0.01} & \textcolor{purple} {0.0} & \textcolor{purple} {-94.73} & \textcolor{purple} {-86.87 } \\

& \textcolor{purple} {7.07} & \textcolor{purple} {7.08} & \textcolor{purple} {50.85} & \textcolor{purple} {49.38} & \textcolor{purple} {10,046.43} & \textcolor{purple} {9,880.93} & \textcolor{purple} {2.0} & \textcolor{purple} {1.96} & \textcolor{purple} {0.0} & \textcolor{purple} {0.02} & \textcolor{purple} {0.0} & \textcolor{purple} {0.0} & \textcolor{purple} {-98.11} & \textcolor{purple} {-96.5 } \\

{\multirow{3}{*}{100 \%}} &  \textcolor{purple}{99.0} & {93.26} & \textcolor{purple} {100.0} & {100.0} & \textcolor{purple} {514,687.26} & {472,315.05} & \textcolor{purple} {51.47} & {47.23} & \textcolor{purple} {0.0} & {3.21} & \textcolor{purple} {0.35} & {0.01} & \textcolor{purple} {-64.97} & {222.34 } \\

& \textcolor{purple} {2.68} & \textcolor{purple} {2.6} & \textcolor{purple} {100.0} & \textcolor{purple} {100.0} & \textcolor{purple} {11,725.34} & \textcolor{purple} {11,529.51} & \textcolor{purple} {1.17} & \textcolor{purple} {1.15} & \textcolor{purple} {0.0} & \textcolor{purple} {0.02} & \textcolor{purple} {0.01} & \textcolor{purple} {0.0} & \textcolor{purple} {-99.42} & \textcolor{purple} {-97.86 } \\

& \textcolor{purple} {2.23} & \textcolor{purple} {2.21} & \textcolor{purple} {100.0} & \textcolor{purple} {100.0} & \textcolor{purple} {10,074.25} & \textcolor{purple} {10,036.04} & \textcolor{purple} {1.01} & \textcolor{purple} {1.0} & \textcolor{purple} {0.0} & \textcolor{purple} {0.01} & \textcolor{purple} {0.0} & \textcolor{purple} {0.0} & \textcolor{purple} {-99.58} & \textcolor{purple} {-99.2 } \\

\bottomrule
\multicolumn{5}{|c|}{\textit{*i.c. = initial collateral $Q_0$}} & \multicolumn{5}{|c|}{\textit{No With. = No Premium Withdrawn}} & \multicolumn{5}{|c|}{\textit{With. = Premium Withdrawn}} \\ 
\bottomrule
\end{tabularx}
\end{table}
\vspace*{-5mm}

\newpage

\newcolumntype{P}[1]{>{\centering\arraybackslash}p{#1}}
\begin{table}[H]

\caption{rkAMM results. Every 90 days a 50\% of the premium obtained is withdrawn. \label{tab:rkAMM_default_results_90days}}
\begin{tabularx}{\linewidth }
{|P{1.3cm}|P{1.3cm}|P{1.3cm}|P{1.3cm}|P{1.3cm}|P{1.6cm}|P{1.6cm}|P{1.3cm}|P{1.3cm}|P{1.5cm}|P{1.3cm}|P{1.3cm}|P{1.3cm}|P{1.35cm}|P{1.35cm}|}

        \toprule
        Scenario & \multicolumn{2}{c|}{\% accepted invoices}& \multicolumn{2}{c|}{\% unpaid invoices}&  \multicolumn{2}{c|}{Avg. loss due to} & \multicolumn{2}{c|}{Total collateral covered} & \multicolumn{2}{c|}{Total premium with.}& \multicolumn{2}{c|}{Remaining premium} & \multicolumn{2}{c|}{AMM profit} \\
            
        \# & \multicolumn{2}{c|}{(collateralized)} & \multicolumn{2}{c|}{(capital not returned)}&             \multicolumn{2}{c|}{unpaid invoices} & \multicolumn{2}{c|}{x (i.c.)}&             \multicolumn{2}{c|}{x (i.c.)}& \multicolumn{2}{c|}{x (i.c.)}& \multicolumn{2}{c|}{percentage} \\
        \toprule

        & No  With. &  With. & No With. & With. & No With. & With. & No With. & With. & No With. & With. & No With. & With. & No With. & With. \\
        \toprule
        
1.1 & 69.96 & 70.01 & 0.0 & 0.0 & 0.0 & 0.0 & 32.73 & 32.71 & 0.0 & 1.04 & 1.1 & 0.18 & 881.2 & 889.71 \\ 
1.2 & 86.84 & 86.55 & 0.0 & 0.0 & 0.0 & 0.0 & 43.56 & 43.39 & 0.0 & 3.56 & 3.93 & 0.45 & 1,591.24 & 1,607.24 \\ 
1.3 & 93.64 & 94.02 & 0.0 & 0.0 & 0.0 & 0.0 & 48.16 & 48.19 & 0.0 & 5.16 & 5.47 & 0.5 & 2,409.92 & 2,435.5 \\ 
1.4 & 99.21 & 99.22 & 0.0 & 0.0 & 0.0 & 0.0 & 51.9 & 51.77 & 0.0 & 6.32 & 6.83 & 0.51 & 4,844.95 & 4,838.42 \\ 
2.1 & 56.46 & 52.61 & 1.95 & 1.95 & 4,693.43 & 4,125.03 & 24.79 & 22.42 & 0.0 & 0.34 & 0.31 & 0.04 & 571.75 & 523.22 \\ 
2.2 & 51.15 & 51.48 & 5.08 & 4.88 & 10,905.05 & 10,265.56 & 21.45 & 21.71 & 0.0 & 0.3 & 0.17 & 0.04 & 436.98 & 451.74 \\ 
2.3 & 34.42 & 34.65 & 20.42 & 20.04 & 24,583.38 & 24,005.7 & 12.07 & 12.05 & 0.0 & 0.14 & 0.05 & 0.01 & 102.84 & 100.99 \\ 
3.1 & 81.62 & 80.54 & 0.0 & 0.0 & 0.0 & 0.0 & 40.82 & 39.86 & 0.0 & 2.65 & 3.01 & 0.38 & 766.99 & 768.17 \\ 
3.2 & 52.74 & 50.55 & 0.0 & 0.0 & 0.0 & 0.0 & 23.44 & 22.15 & 0.0 & 0.39 & 0.33 & 0.07 & 566.55 & 543.55 \\ 
3.3 & 32.17 & 31.62 & 0.0 & 0.0 & 0.0 & 0.0 & 12.71 & 12.48 & 0.0 & 0.13 & 0.03 & 0.01 & 332.73 & 323.79 \\ 
4.1 & 100.0 & 100.0 & 0.0 & 0.0 & 0.0 & 0.0 & 5.0 & 5.0 & 0.0 & 0.69 & 0.73 & 0.05 & 72.64 & 73.59 \\ 
4.2 & 49.71 & 46.7 & 0.0 & 0.0 & 0.0 & 0.0 & 24.86 & 23.35 & 0.0 & 0.41 & 0.31 & 0.06 & 590.31 & 560.45 \\ 
4.3 & 22.91 & 20.52 & 0.0 & 0.0 & 0.0 & 0.0 & 28.64 & 25.65 & 0.0 & 0.63 & 0.21 & 0.06 & 762.75 & 700.41 \\ 
5.1 & 96.81 & 96.73 & 0.0 & 0.0 & 0.0 & 0.0 & 50.16 & 50.17 & 0.0 & 12.5 & 13.51 & 1.24 & 2,177.66 & 2,199.24 \\ 
5.2 & 32.29 & 31.89 & 0.0 & 0.0 & 0.0 & 0.0 & 10.79 & 10.58 & 0.0 & 0.09 & 0.03 & 0.01 & 113.95 & 112.06 \\ 
5.3 & 23.82 & 23.33 & 0.0 & 0.0 & 0.0 & 0.0 & 7.16 & 7.1 & 0.0 & 0.02 & 0.01 & 0.0 & 8.54 & 8.48 \\ 

\bottomrule
\multicolumn{5}{|c|}{\textit{*i.c. = initial collateral $Q_0$}} & \multicolumn{5}{|c|}{\textit{No With. = No Premium Withdrawn}} & \multicolumn{5}{|c|}{\textit{With. = Premium Withdrawn}} \\ 
\bottomrule
\end{tabularx}
\end{table}

\begin{table}[H]
\caption{rkAMM hack scenario results. Every 90 days a 50\% of the premium obtained is withdrawn.\label{tab:rkAMM_hack_results_90days}}
\begin{tabularx}{\linewidth }
{|P{1.3cm}|P{1.3cm}|P{1.3cm}|P{1.3cm}|P{1.3cm}|P{1.635cm}|P{1.635cm}|P{1.3cm}|P{1.3cm}|P{1.5cm}|P{1.3cm}|P{1.3cm}|P{1.3cm}|P{1.35cm}|P{1.35cm}|}

        \toprule
        Hack & \multicolumn{2}{c|}{\% accepted invoices}& \multicolumn{2}{c|}{\% unpaid invoices}&  \multicolumn{2}{c|}{Avg. loss due to} & \multicolumn{2}{c|}{Total collateral covered} & \multicolumn{2}{c|}{Total premium with.}& \multicolumn{2}{c|}{Remaining premium} & \multicolumn{2}{c|}{AMM profit} \\
            
        Prob. & \multicolumn{2}{c|}{(collateralized} & \multicolumn{2}{c|}{(collateralized)}&             \multicolumn{2}{c|}{unpaid invoices} & \multicolumn{2}{c|}{x (i.c.)}&             \multicolumn{2}{c|}{x (i.c.)}& \multicolumn{2}{c|}{x (i.c.)}& \multicolumn{2}{c|}{percentage} \\
        \toprule

        & No  With. &  With. & No With. & With. & No With. & With. & No With. & With. & No With. & With. & No With. & With. & No With. & With. \\
        \toprule

{\multirow{3}{*}{10 \%}} &  99.96 & 99.95 & 10.08 & 9.87 & 52,195.52 & 51,985.91 & 52.44 & 52.4 & 0.0 & 15.42 & 16.39 & 1.34 & 2,301.21 & 2,337.35 \\ 

& 33.05 & 32.1 & 10.08 & 10.04 & 10,762.52 & 10,505.58 & 10.83 & 10.45 & 0.0 & 0.12 & 0.04 & 0.01 & 85.23 & 80.5 \\

& \textcolor{purple}{ 19.02 } & \textcolor{purple}{18.4 } & \textcolor{purple}{10.0 } & \textcolor{purple}{10.07 } & \textcolor{purple}{5,192.23 } & \textcolor{purple}{5,353.66 } & \textcolor{purple}{5.29 } & \textcolor{purple}{5.16 } & \textcolor{purple}{0.0 } & \textcolor{purple}{0.02 } & \textcolor{purple}{0.01 } & \textcolor{purple}{0.0 } & \textcolor{purple}{-45.62 } & \textcolor{purple}{-47.39 } \\ 

{\multirow{3}{*}{50 \%}} & { 99.89 } & {99.91 } & {50.12 } & {49.94 } & {262,371.99 } & {260,757.48 } & {52.29 } & {52.35 } & {0.0 } & {3.25 } & {1.79 } & {0.25 } & {483.09 } & {659.2 } \\

& \textcolor{purple}{ 11.59 } & \textcolor{purple}{10.17 } & \textcolor{purple}{49.47 } & \textcolor{purple}{50.4 } & \textcolor{purple}{14,717.05 } & \textcolor{purple}{13,754.29 } & \textcolor{purple}{3.06 } & \textcolor{purple}{2.69 } & \textcolor{purple}{0.0 } & \textcolor{purple}{0.05 } & \textcolor{purple}{0.01 } & \textcolor{purple}{0.0 } & \textcolor{purple}{-94.57 } & \textcolor{purple}{-91.64 } \\ 

& { 7.21 } & \textcolor{purple}{7.23 } & \textcolor{purple}{50.1 } & \textcolor{purple}{49.61 } & \textcolor{purple}{10,057.43 } & \textcolor{purple}{9,962.06 } & \textcolor{purple}{2.02 } & \textcolor{purple}{2.02 } & \textcolor{purple}{0.0 } & \textcolor{purple}{0.01 } & \textcolor{purple}{0.0 } & \textcolor{purple}{0.0 } & \textcolor{purple}{-98.19 } & \textcolor{purple}{-97.24 } \\ 

{\multirow{3}{*}{100 \%}} & \textcolor{purple}{ 99.08 } & {97.19 } & \textcolor{purple}{100.0 } & {100.0 } & \textcolor{purple}{518,522.18 } & {501,747.51 } & \textcolor{purple}{51.85 } & {50.17 } & \textcolor{purple}{0.0 } & {1.18 } & \textcolor{purple}{0.39 } & {0.09 } & \textcolor{purple}{-60.73 } & {26.59 } \\ 

& \textcolor{purple}{ 2.71 } & \textcolor{purple}{2.66 } & \textcolor{purple}{100.0 } & \textcolor{purple}{100.0 } & \textcolor{purple}{11,712.08 } & \textcolor{purple}{11,620.56 } & \textcolor{purple}{1.17 } & \textcolor{purple}{1.16 } & \textcolor{purple}{0.0 } & \textcolor{purple}{0.01 } & \textcolor{purple}{0.01 } & \textcolor{purple}{0.0 } & \textcolor{purple}{-99.38 } & \textcolor{purple}{-98.56 } \\ 

& \textcolor{purple}{ 2.23 } & \textcolor{purple}{2.19 } & \textcolor{purple}{100.0 } & \textcolor{purple}{100.0 } & \textcolor{purple}{10,076.63 } & \textcolor{purple}{10,054.62 } & \textcolor{purple}{1.01 } & \textcolor{purple}{1.01 } & \textcolor{purple}{0.0 } & \textcolor{purple}{0.01 } & \textcolor{purple}{0.0 } & \textcolor{purple}{0.0 } & \textcolor{purple}{-99.61 } & \textcolor{purple}{-99.39 } \\

\bottomrule
\multicolumn{5}{|c|}{\textit{*i.c. = initial collateral $Q_0$}} & \multicolumn{5}{|c|}{\textit{No With. = No Premium Withdrawn}} & \multicolumn{5}{|c|}{\textit{With. = Premium Withdrawn}} \\ 
\bottomrule
\end{tabularx}
\end{table}

\end{landscape}

\restoregeometry

\newgeometry{left=1.5cm,right=1.5cm,top=0.3cm,bottom=0.3cm}
\begin{landscape}

\begin{figure}[H]
        \centering
        \vfill
        \includegraphics[scale=0.47]{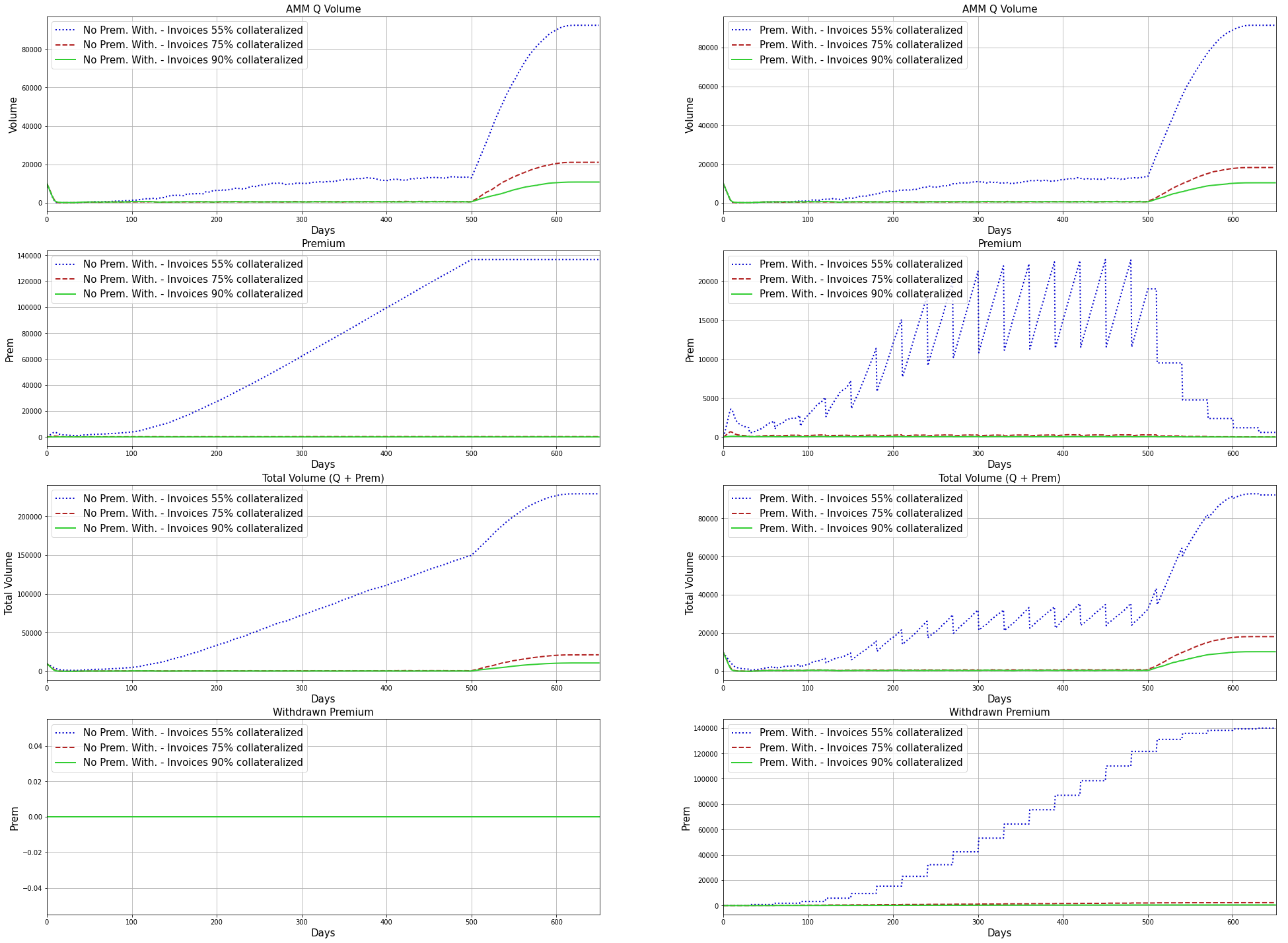}
        \vfill
\caption{AMM Curves from simulation of scenario 5 and 30 days of withdrawal period.\label{scenario_5_all_curves_30days}}
\end{figure}

\end{landscape}
\restoregeometry

\newpage

In addition to the different scenarios results from tables presented above, in following Figures \ref{prem_no_prem_AMM_profit} and \ref{prem_no_prem_diff_AMM_profit} a comparative analysis of the AMM profit of all the scenarios can be observed. Figure \ref{prem_no_prem_AMM_profit} shows a comparison between the AMM profit  when there is no premium withdrawal and when there is, depending on the period in which the premium is withdrawn. Figure \ref{prem_no_prem_diff_AMM_profit} shows the AMM profit percentage difference   between scenarios where the premium is withdrawn and where it is not. Premium withdrawal intervals are 1,30 and 90 days.

\begin{figure}[H]
\centering
\includegraphics[width=.75\textwidth]{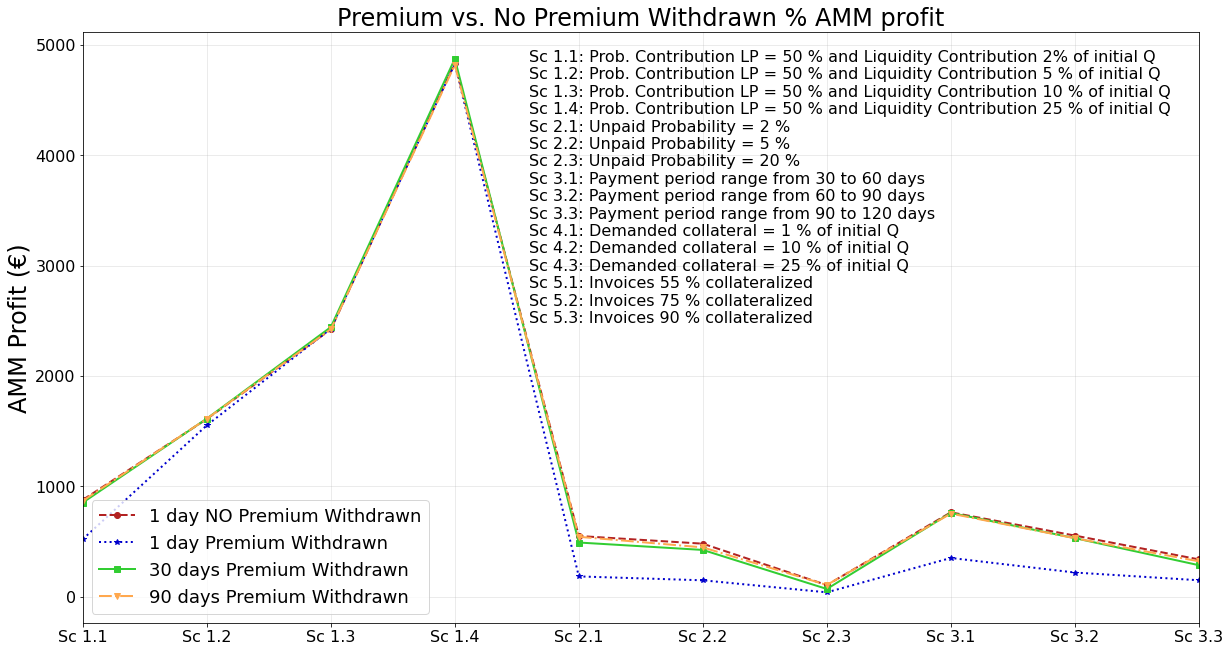}
\caption{Premium vs. No Premium Withdrawn AMM profit absolute values from scenario 5.\label{prem_no_prem_AMM_profit}}
\end{figure}   

\vspace{-5mm}

\begin{figure}[H]
\centering
\includegraphics[width=.75\textwidth]{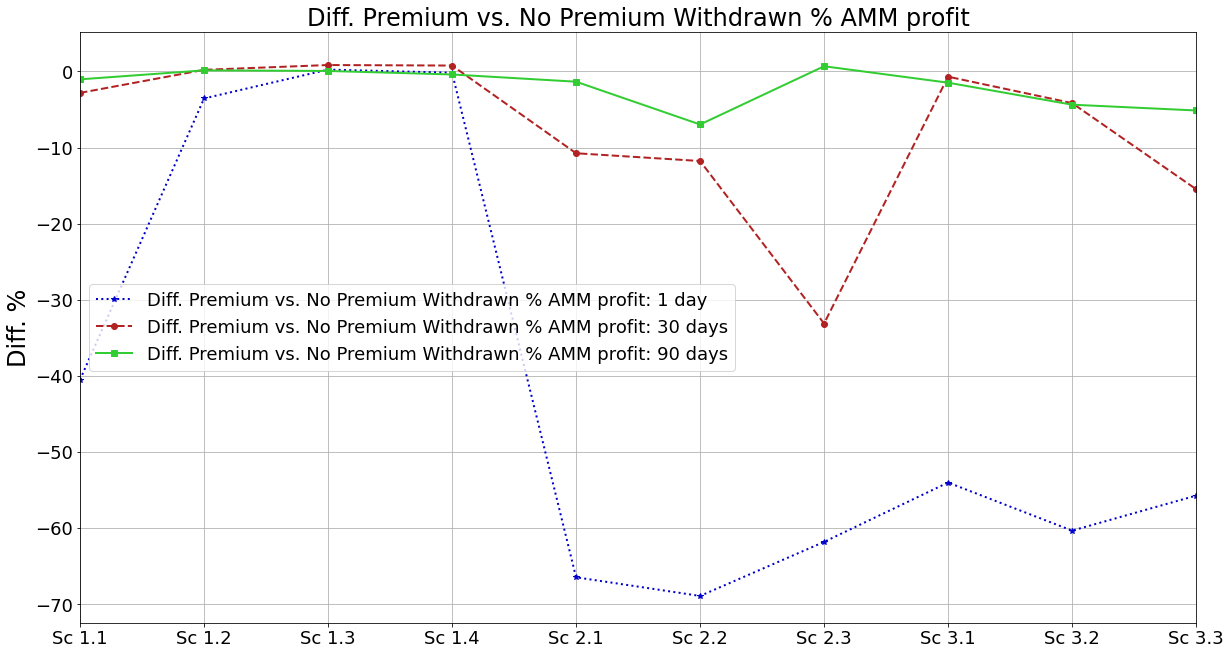}
\caption{Premium vs. No Premium Withdrawn AMM profit difference from scenario 5.\label{prem_no_prem_diff_AMM_profit}}
\end{figure}   
\vspace{-5mm}

As can be seen, the AMM profit is not affected by the premium withdrawal in the majority of the analyzed scenarios. Also, it can even be concluded that the period in which the premium is withdrawn does not have a great impact either. However, in scenarios where the premium is withdrawn every day, the performance of the AMM does suffer and the results obtained in this case are worse.

Furthermore, although in some scenarios the daily premium withdrawal causes the AMM profit to be only a few points below the AMM profit when there is no premium withdrawal, in other scenarios the AMM profit obtained is almost 70\% lower when there is premium withdrawal.


\newgeometry{left=2.5cm,top=0.4cm,bottom=0.3cm}
\begin{landscape}

\begin{figure}
        \centering\vfill
        \includegraphics[scale=0.25]{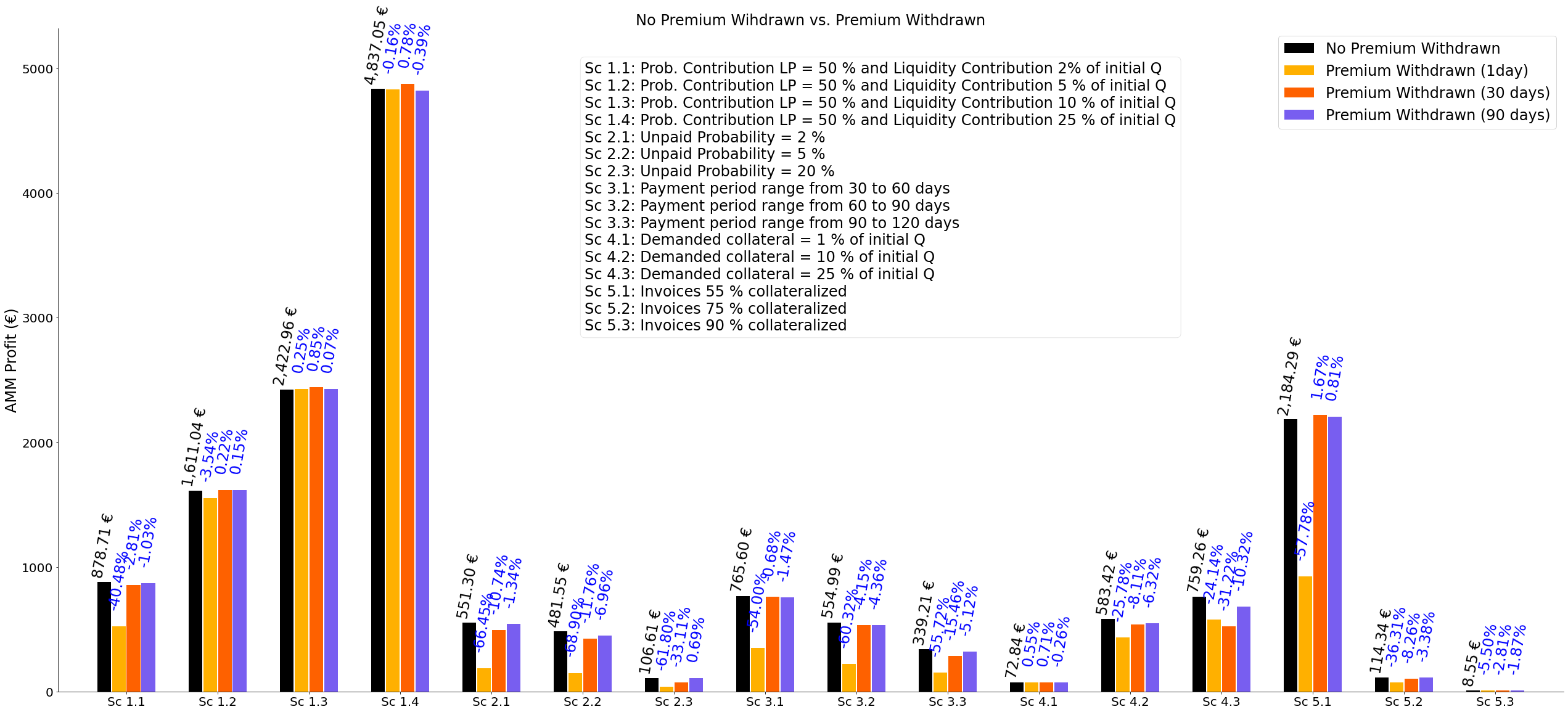}
        \vfill
        \caption{No Premium Withdrawn vs. Premium Withdrawn.
        \label{prem_vs_no_prem_global}}
\end{figure}

\end{landscape}
\restoregeometry

\newpage



\newgeometry{left=1.5cm,right=1.5cm,top=0.3cm,bottom=0.3cm}
\begin{landscape}

\begin{figure}[H]
        \centering
        \vfill
        \includegraphics[scale=0.47]{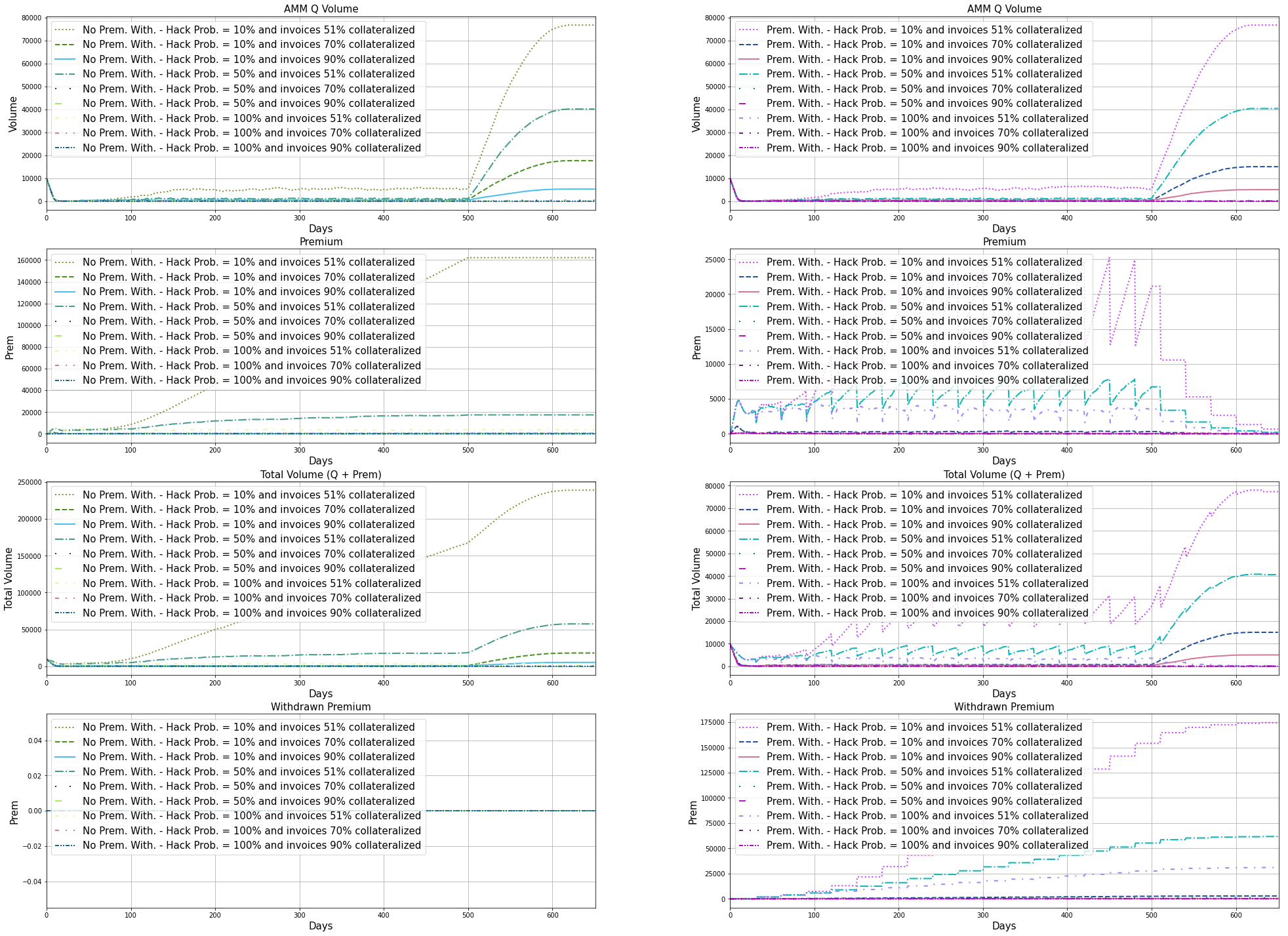}
        \vfill
\caption{AMM Curves from simulation of hack scenario and 30 days of withdrawal period.\label{hack_scenario_all_curves_30days}}
\end{figure}

\end{landscape}
\restoregeometry

\newpage

\begin{figure}[H]
\centering
\includegraphics[width=.76\textwidth]{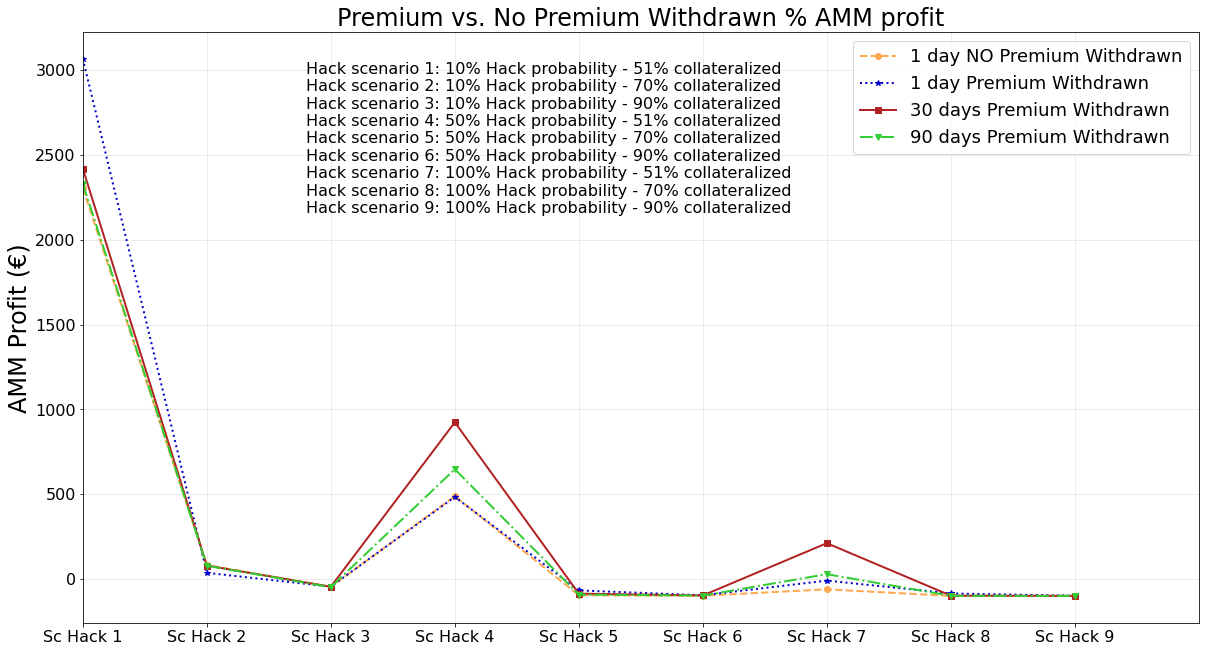}
\caption{Premium vs. No Premium Withdrawn AMM profit absolute values from hack scenario.\label{hack_prem_no_prem_AMM_profit}}
\end{figure}   


From Figure \ref{hack_prem_no_prem_AMM_profit} it can be seen that practically a third of the scenarios is affected to a greater extent by the premium withdrawal, and therefore the AMM profit is altered. Further, this impact turns out to be positive and the performance of the AMM for this set of scenarios is better when premium is withdrawn periodically, especially with invoices collateralized at 51\%. The best cases are those in which the invoices are collateralized at 51\%, although the AMM profit decreases as the hack probability increases.

\begin{figure}[H]
\centering
\includegraphics[width=.76\textwidth]{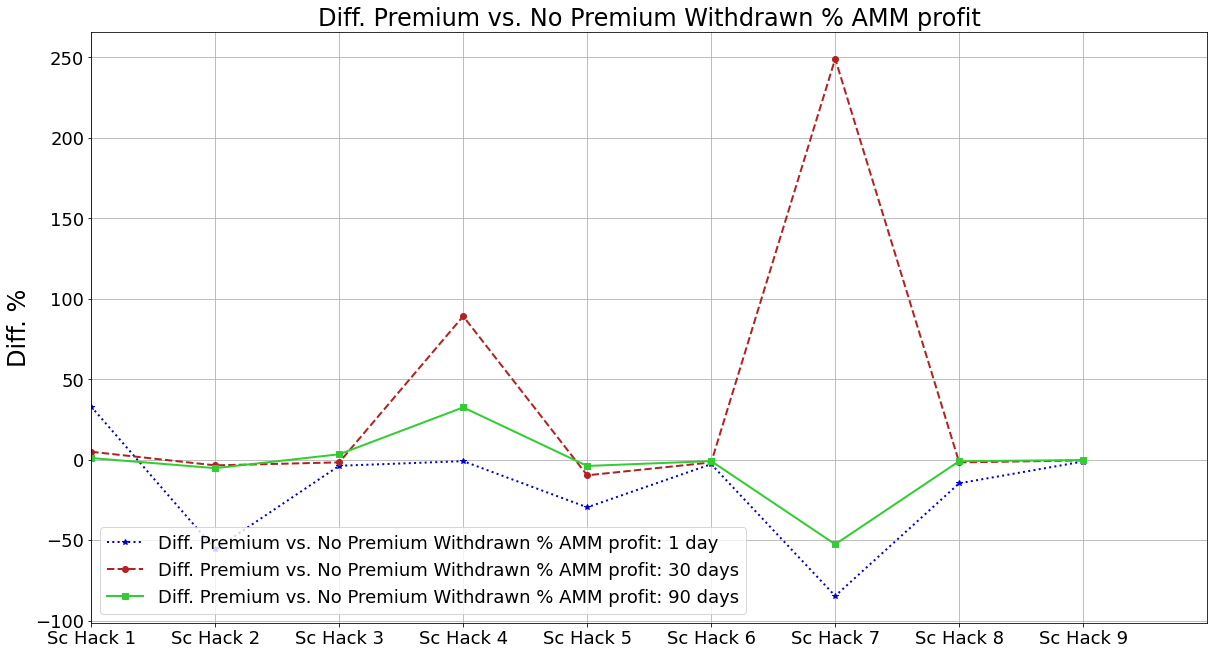}
\caption{Premium vs. No Premium Withdrawn AMM profit difference from hack scenario.\label{hack_prem_no_prem_diff_AMM_profit}}
\end{figure}   
\vspace{-5mm}

As for the percentage difference with respect to withdrawing premium or not shown in Figure \ref{hack_prem_no_prem_diff_AMM_profit}, the scenarios where premium is withdrawn every 30 days with 51\% collateralized invoices present an improvement of 5\%, 89\% and 249\% with hacking probabilities of 10\%, 50\% and 100\% respectively. Regarding daily withdrawals, the only favorable case is found when invoices are collateralized at 51\% with a hacking probability of 10\%. Finally, for premium withdrawals every 90 days, the most remarkable case is in which invoices are collateralized at 51\% with a hack probability of 50\%. The rest of the scenarios have presented a worse performance when there ir a periodic premium withdrawal.

\newgeometry{left=2.5cm,top=0.4cm,bottom=0.3cm}
\begin{landscape}

\begin{figure}
        \centering\vfill
        \includegraphics[scale=0.25]{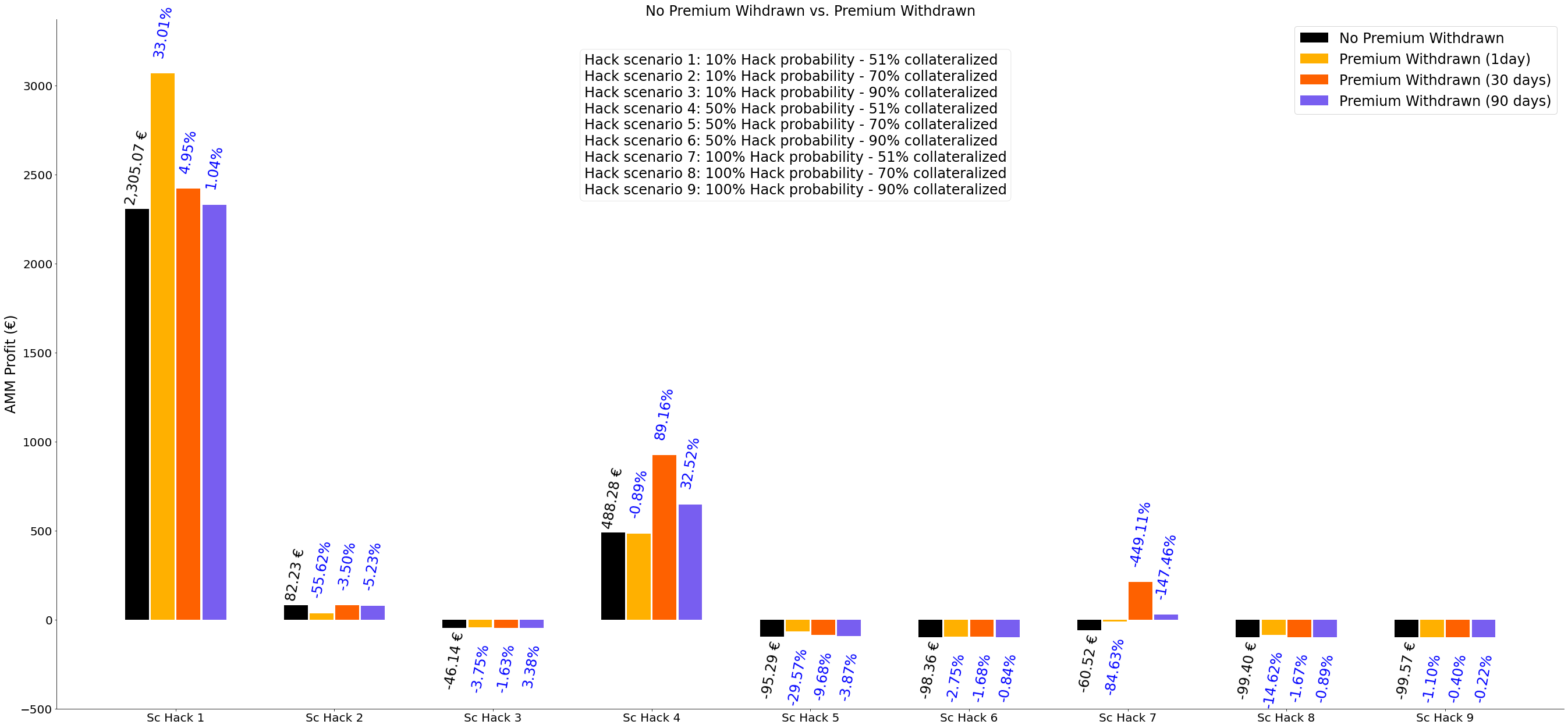}
        \vfill
        \caption{No Premium Withdrawn vs. Premium Withdrawn. (Hack scenarios)
        \label{hack_prem_vs_no_prem_global}}
\end{figure}

\end{landscape}
\restoregeometry

\newpage



\section{Withdrawn Premium Analysis}

In this section, a detailed analysis of the results obtained and presented in the previous section is carried out.

First, it must be taken into account that the AMM compensates invoices from the premium if it does not have sufficient liquidity. This happens for all scenarios and affects both the final premium values and the total volume of the AMM.

Considering the results obtained, the more adverse the scenario, the lower the premium withdrawn from the AMM will be. The same happens with the remaining premium in the AMM, the worse the scenario, the less amount will be available. This is because the AMM will try to collateralize with the premium due to lack of liquidity, which is the expected behaviour in adverse scenarios.

\textit{Scenario 1: The greater the increase of liquidity, it is almost possible to match the results of this scenario when no premium is withdrawn.}

As liquidity increases, the difference from an approximate -50\% to a -1\% difference on accepted/paid invoices increases. In all cases, all collateralized invoices are paid.

When there is a greater liquidity contribution, the total covered is increased. However, in the worst case of liquidity contribution (scenario 1.1) there is a 60\% difference in terms of the total collateral covered. This difference is drastically reduced in the 1.4 scenario where it is only 1\%.

Regarding the premium managed by the AMM in these cases. It is increasing at the same time that greater liquidity is provided, which was to be expected given that a greater number of invoices are collateralized.

Regarding the benefit of the AMM, the greater the contribution of liquidity, the greater the profit AMM. However, the withdrawal of premium does not affect the benefit of the premium by itself, since there are scenarios in which premium is withdrawn and a greater benefit is obtained (even if only 1\%) than those in which there is no withdrawal of premium. This is because in scenarios where there is no premium withdrawal, less premium is collected than in cases where premium is withdrawn.

\textit{Scenario 2: Faced with an increase in unpaid invoices, the relationship between the scenarios with premium withdrawal and those without it is maintained, although the absolute values are affected and the results are worse the higher the number of unpaid invoices there are.}

The number of collateralized invoices decreases, but the relationship between the scenarios with premium withdrawal and those without it is maintained. The same happens for the amount of collateral covered and the AMM profit.

Regarding the losses due to unpaid invoices, the relationship between scenarios is maintained despite the increase in the number of these invoices.

\textit{Scenario 3: The greater the delay in the payment of the invoices, the lower the premium that can be withdrawn and the lower the remaining premium in the case where the premium is not withdrawn. The shorter the payment term, the better the liquidity of the rkAMM.}

The number of collateralized invoices will be less, but the relationship between the scenarios with premium withdrawal and those without is variable, and it worsens in the scenarios where there are greater delays in the invoice payments. The same happens for the amount of collateral covered.

The AMM profit is lower as the delay in the payment of invoices increases. However, this is not directly related to the relationship between the scenarios with premium withdrawal and those without it. In fact, it can be verified that the scenario with the greatest decompensation is 3.2, where the delay in payments is not the worst of all the possible ones.

\textit{Scenario 4: The greater the amount to be collateralized, the better the results of the AMM are in terms of its performance since more liquidity is collateralized. Therefore, the higher premium is obtained and the greater the benefit of the AMM. In summary, better results are achieved the higher the risk the rkAMM assumes.}

The number of collateralized invoices decreases, but the relationship between the scenarios with premium withdrawal and those without is variable, and it worsens in the scenarios where there are increasingly large invoices. Despite the fact that the collateral covered and the AMM profit increase, the relationship between scenarios has the same behavior.

\textit{Scenario 5: Same pattern as scenario 4. The higher the percentage to be collateralized, the worse the results of the AMM in terms of its performance, and it can be determined that the higher the percentage of collateralization of the invoices, the more similar the results are between scenarios where premium is withdrawn and where it is not.}

The number of collateralized invoices decreases, however, the relationship between the scenarios in which premium is withdrawn and those that are not, decreases until it is almost equal to the scenario where the invoices are almost completely collateralized. The same happens for the amount of collateral covered and profit AMM as the amount to be collateralized increases.

\textit{Hack Scenario: The greater the hack probability and the amout to collateralize, the worse the results of the AMM in terms of its performance. However, in some cases, premium withdrawal performs better in terms of AMM profit for withdrawal periods of 30 and 90 days.}

Losses due to unpaid invoices, as expected, increase as the probability of hacking increases. On the contrary, the amount of collateral covered decreases as the probability of hacking and the amount to be collateralized increases. The same happens for the profit AMM.

Finally, as a conclusion to this section, the following comments are made.
\begin{itemize}
    \item Distributing the premium can give better results in scenarios with greater liquidity (those with less payment terms or more liquidity contribution) and with higher risk (fewer guarantees or greater defaulting).
    \item Thus, we find ourselves with an "aseptic" measure, withdrawing the benefits to improve the performance of rkAMM. The policy of giving out premium affects the design and performance of the rkAMM, and the following could be recommended:
    \begin{enumerate}
        \item Not sharing profits is not an adequate policy for the operation of the rkAMM.
        \item Make a variable policy based on liquidity, defaulting and payment terms.     
    \end{enumerate}
    \item Apart from that, distributing benefits early is expected to have a bandwagon effect on the demand for rkAMM products with more and more liquidity contributors.

\end{itemize}


\section{Discussion}


The following considerations and comments are made on the different scenarios analyzed over the results of previous sections about the impact on the number of accepted invoices, the number of unpaid invoices, the average losses, collateral covered, remaining and withdrawn premium, and finally some comments on the final profit achieved. The comments below do not differentiate between premium withdrawal or not, but are analyzed together. In fact, it is important to highlight that both the cases where premium is withdrawn and in those which are not, the same behavior trend exists between the scenarios.

The rkAMM will be able to accept more invoices provided that the liquidity stream flows in the LP with constant contribution probability (as in scenario 1). Otherwise, if there is no liquidity stream then the rkAMM gets its liquidity depleted progressively, then the rkAMM continues collateralizing invoices using the remaining premium until liquidity is replenished with premium or invoices are repaid. Also, it may happen that the premium runs out and invoices cannot be collateralized, in which case liquidity in rkAMM will be expected to be replenished. Scenarios with small invoices in terms of amount to be collateralized and small high percentage to be collateralized present the best results when it comes to the number of accepted invoices.

As for losses due to unpaid invoices, only scenario 2 shows values above zero. This result was expected since unpaid invoices are intentionally included to be processed by the rkAMM. The hack scenarios, logically, present losses because it is an input condition of these cases to be able to be analyzed. The more unpaid invoices, the greater the losses, however these losses increase considerably when the invoice amount to be collateralized is greater. 

Continuing with the total collateral covered, the highest values of collateral covered are achieved in those scenarios where there is a liquidity contribution by the LPs, this allows the rkAMM to have margin to cover more collateral. In the rest of the scenarios, as a general rule, the values of collateral covered obtained do not reach the values of scenario 1. However, it is worth mentioning that similar values of total collateral covered are achieve for scenarios 3.1 and 5.1 when there is no premium  withdrawn. 

Regarding the premium withdrawn from the rkAMM when there is premium withdrawal, those scenarios that have a greater total collateral covered will, in turn, have a greater amount of premium withdrawn, as expected. On the other hand, the fact of making premium withdrawals in increasingly longer periods means that the premium withdrawn is lower. For the remaining premium of rkAMM when there is no premium withdrawal, the behavior is analogous. The values obtained in both cases show that the premium is in constant circulation and that it is used for the collateralization of invoices.

Finally, about the rkAMM profit percentage, an analysis of this variable is made in more detail in previous sections. However, it can be concluded that all scenarios, except the hacking ones, achieve a positive profit. And if we make a comparison between scenarios where premium is withdrawn with respect to the ones in which premium is not withdrawn, we can conclude that the scenarios that best adapt and achieve the closest values to the scenarios without premium withdrawal are the ones where premium is withdrawn every 30 days.

In the case of hack scenarios, an analysis of the variables is not carried out since what should be verified is resilience of the rkAMM to an adversary situation. We consider the rkAMM to be resilient when it obtains a positive AMM profit at the end of the simulations. Regardless of the existence of premium withdrawal or not, it is observed that the rkAMM is resilient and can withstand an attack in scenarios with a small hack probability and a low percentage of invoice collateralization, specifically from 49\% to a maximum of 30\%. Although this 30\% limit is considered flexible and with high probability the AMM admits a lower percentage of collateralization such as 20\% to remain resilient. It is important to highlight the scenarios in which the premium is withdrawn every 30 or 90 days with a 100\% probability of hacking when the invoices are 51\% collateralized. In these cases the rkAMM is also resistant.
    
As a conclusion and summary of the hack scenario, the optimal attack is to send bogus invoices with the highest percentage to collateralize allowed, that is, 90\% in our experiments.

We consider them as promising results as there might be mechanisms to resist the hack by checking the invoices, enhance the KYC, and then retain the collateral in custody for doing so or even to impose as penalty the loss of the collateral. We need to investigate further on these mechanisms in a next paper. As well, further research on how to withdraw the profit out of the rkAMM without resenting its operation and long term sustainability must be performed.

\section{Conclusions}


As as summary of the experiments, the optimal operation mode of our rkAMM is with invoices of collateral $p$ over 50\% - 70\%, that proves strong resilience even at scenarios of up to 10\% bogus invoices, with delays of 30 to 120 days and 5\% of nonpayments. In addition, in the event that premium withdrawal exists, it is recommended to use an approach with periodic withdrawals of 30 days.

Under these conditions, there might be room for mechanisms of PoE to check bogus invoices or double factor invoicing on the blockchain to work along our rkAMM, notably the crowd collateralization, the verifiable credentials, the on-chain scoring, or own byppay referral among groups of three actors. 

After experiencing with several hack scenarios, our proposal of the AMM implies that the invoice percentage to collateralize with our rkAMM is advised to be never more than 49\% or lower than 5\%.

\vspace{6pt} 




\section*{Funding}

This research was funded by the grant of  University College London UCL - CBT  3rd Call for Research Proposals 2022 on Distributed Ledger Technologies to the project New Decentralized Compensations of Invoices – ByPay

\section*{Acknowledgments}
This work has been tested thanks to the insights and comments of the Centre de Blockchain de Catalunya CBCat.io and the TECNIO Centre EASY of the University of Girona.




\section*{Abbreviations}
The following abbreviations are used in this manuscript:

\noindent 
\begin{tabular}{@{}ll}

AMM & Automated Market Maker \\
APY & Annual Percentage Yield \\
DeFi & Decentralized Finance \\
DEX & Decentralized Exchange \\
DLT & Distributed Ledger Technology \\
ERP & Enterprise Resource Planning \\
IOU & I Owe You \\
IoV & Internet of Value \\
LP & Liquidity Pool \\
PLF & Protocols for Loanable Funds \\
PoE & Proof of Existence \\
rkAMM & Reverse Kelly Automated Market Maker \\
SME & Small and Medium-sized Enterprise

\end{tabular}


\bibliographystyle{unsrt}  
\bibliography{bibliography}


\appendix

\section[\appendixname~\thesection]{Result curves from 1, 2, 3, 4, 5, and hack scenarios}  \label{appendix}

In this appendix all the curves for the results of all the simulated scenarios can be found. Each subsection corresponds to a certain scenario and its corresponding premium withdrawal period.

\newpage

\newgeometry{left=2cm,right=1.5cm,top=0.3cm,bottom=0.3cm}

\begin{landscape}

\subsection[\appendixname~\thesubsection]{Scenario 1 - 1 day premium withdrawn}
\vfill
\begin{figure}[H]
        \centering
        \includegraphics[scale=0.35]{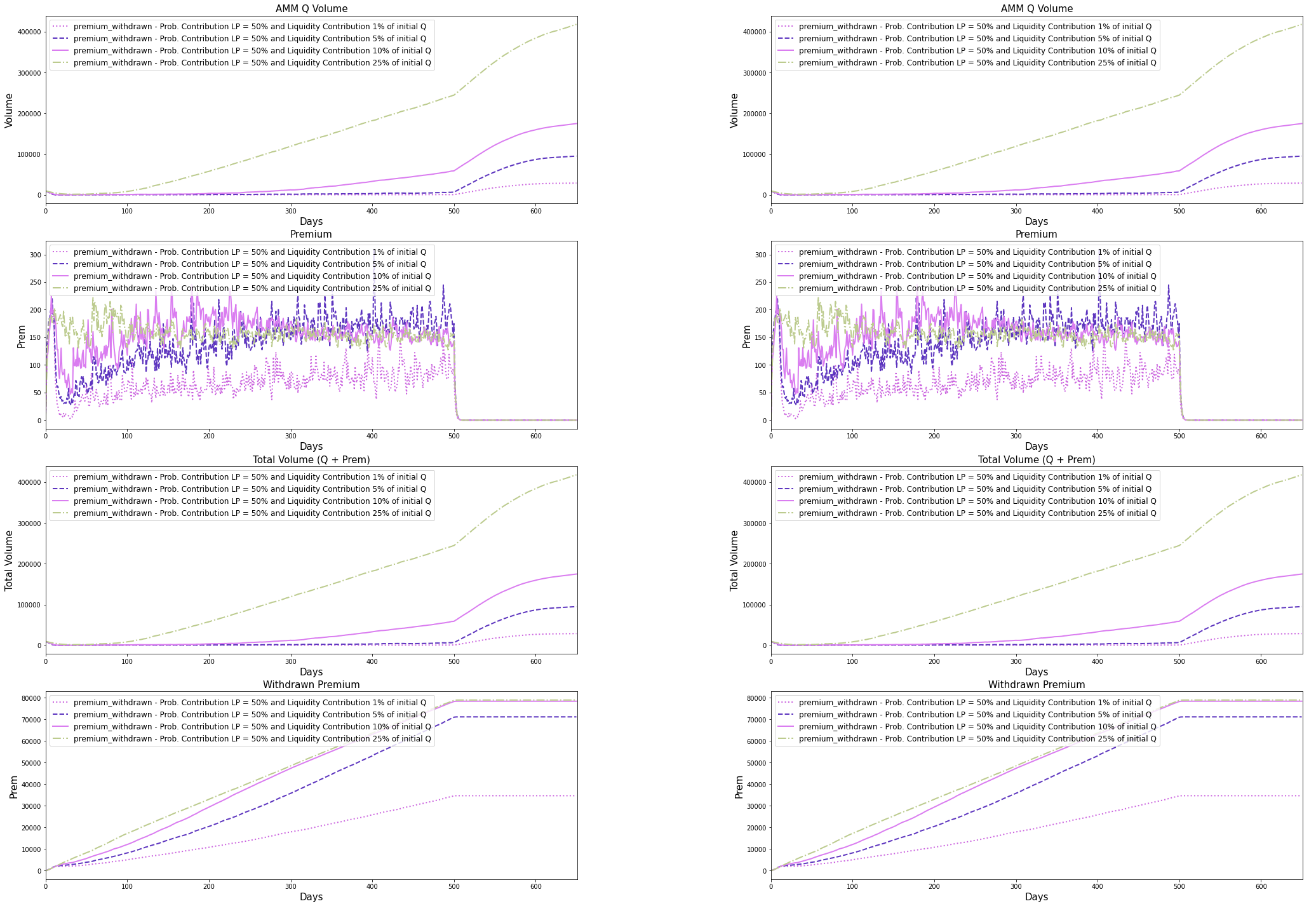}
        \caption{AMM Curves from simulation of scenario 1 and 1 day of withdrawal period.
        \label{scenario_1_all_curves_1day}}
\end{figure}
        \vfill

\end{landscape}

\begin{landscape}

\subsection[\appendixname~\thesubsection]{Scenario 1 - 30 days premium withdrawn}
\vfill
\begin{figure}[H]
        \centering
        \includegraphics[scale=0.35]{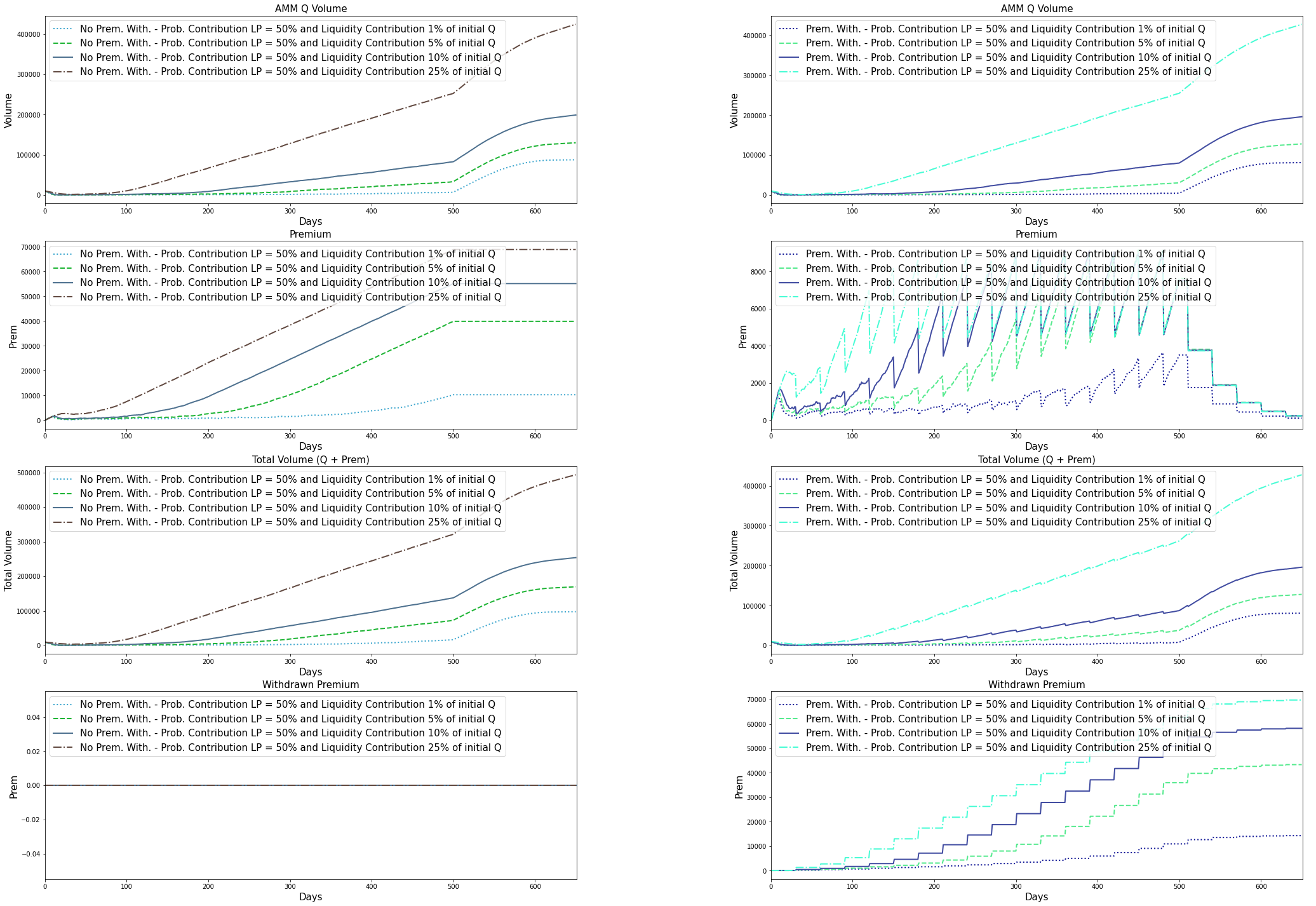}
        \caption{AMM Curves from simulation of scenario 1 and 30 days of withdrawal period.
        \label{scenario_1_all_curves_30days}}
\end{figure}
        \vfill

\end{landscape}

\begin{landscape}

\subsection[\appendixname~\thesubsection]{Scenario 1 - 90 days premium withdrawn}
\vfill
\begin{figure}[H]
        \centering
        \includegraphics[scale=0.35]{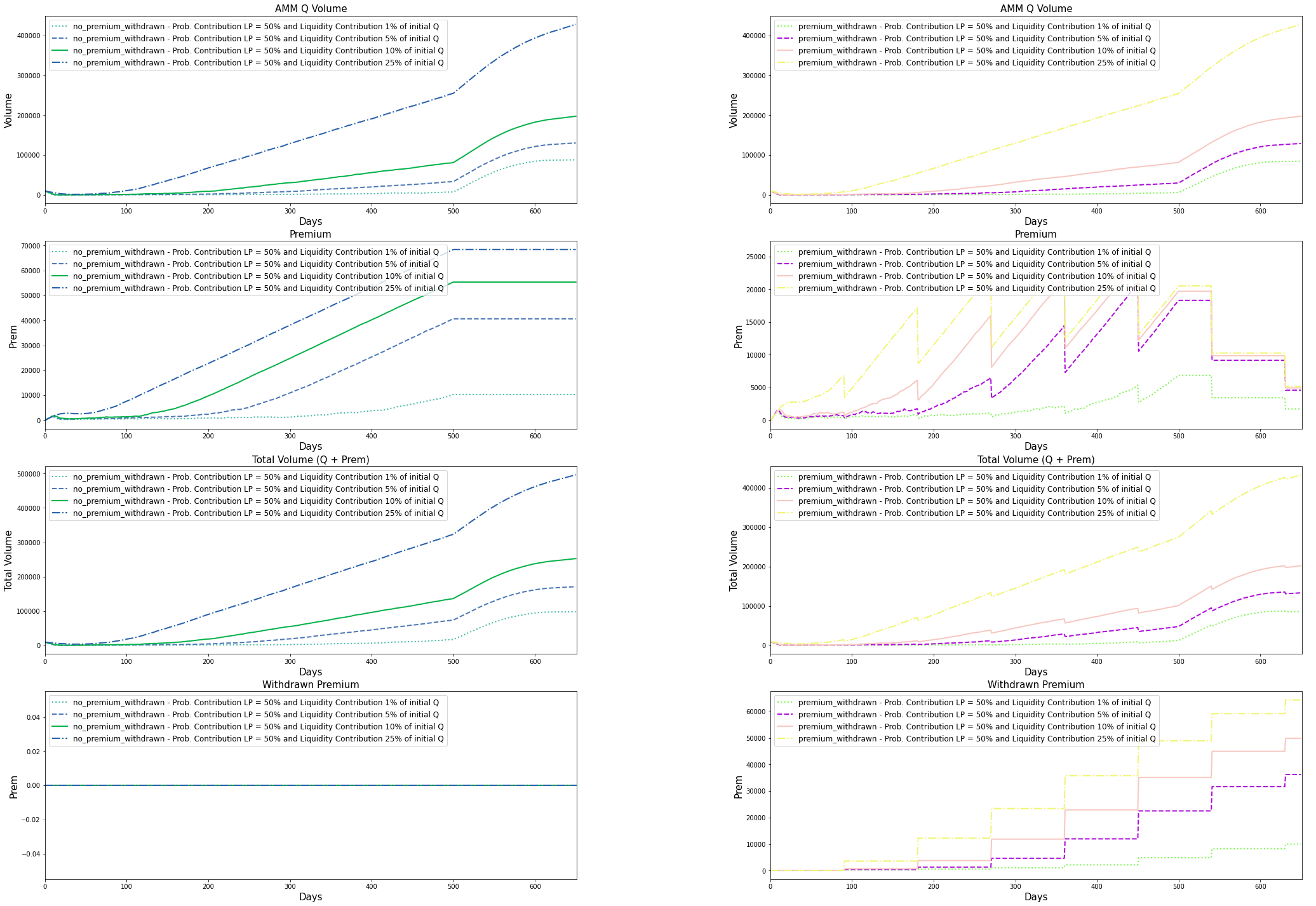}
        \caption{AMM Curves from simulation of scenario 1 and 90 days of withdrawal period.
        \label{scenario_1_all_curves_90days}}
\end{figure}
\vfill

\end{landscape}

\begin{landscape}

\subsection[\appendixname~\thesubsection]{Scenario 2 - 1 day premium withdrawn}
\vfill
\begin{figure}[H]
        \centering
        \includegraphics[scale=0.35]{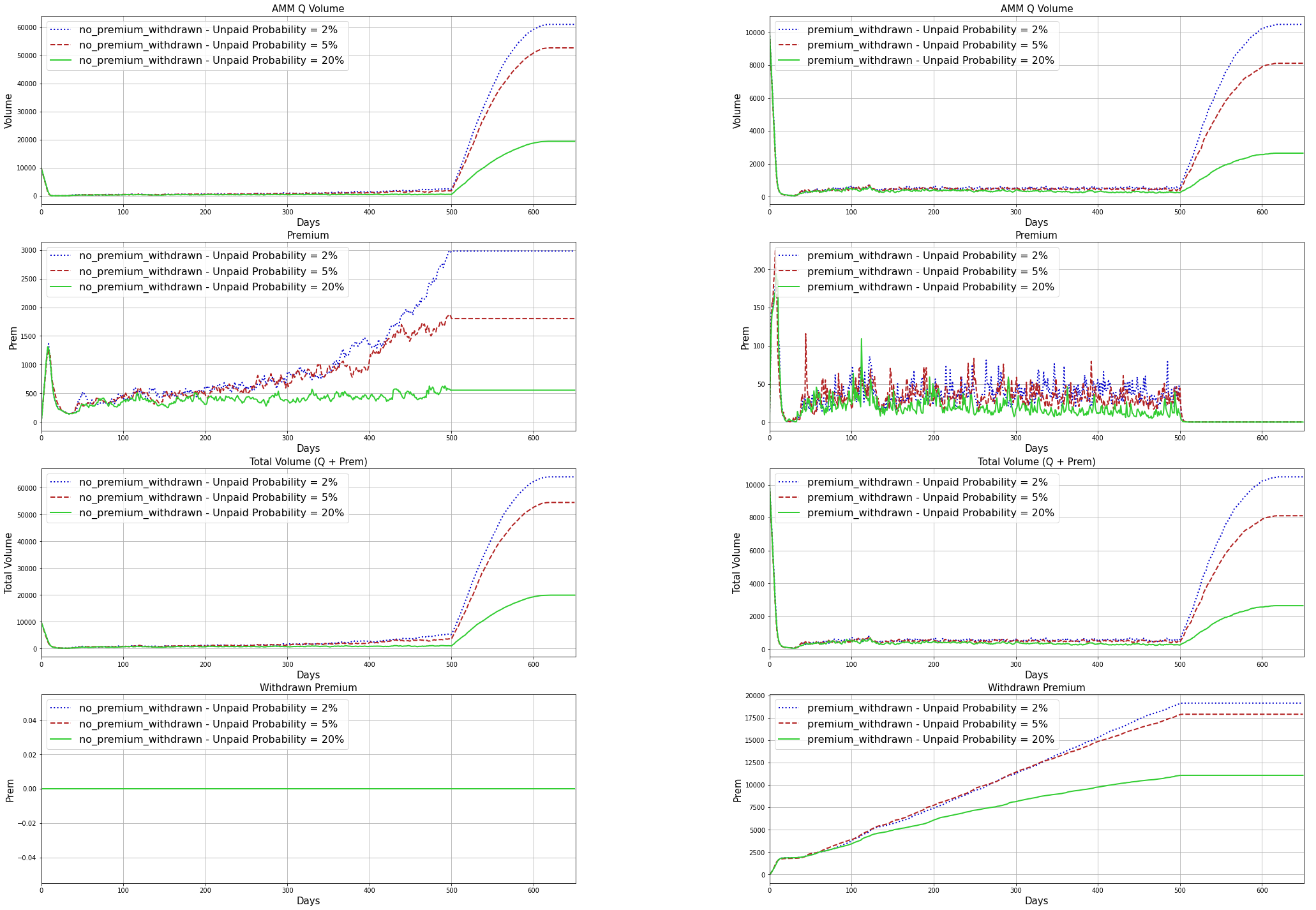}
        \caption{AMM Curves from simulation of scenario 2 and 1 day of withdrawal period.
        \label{scenario_2_all_curves_1day}}
\end{figure}
\vfill

\end{landscape}

\begin{landscape}

\subsection[\appendixname~\thesubsection]{Scenario 2 - 30 days premium withdrawn}
\vfill
\begin{figure}[H]
        \centering
        \includegraphics[scale=0.35]{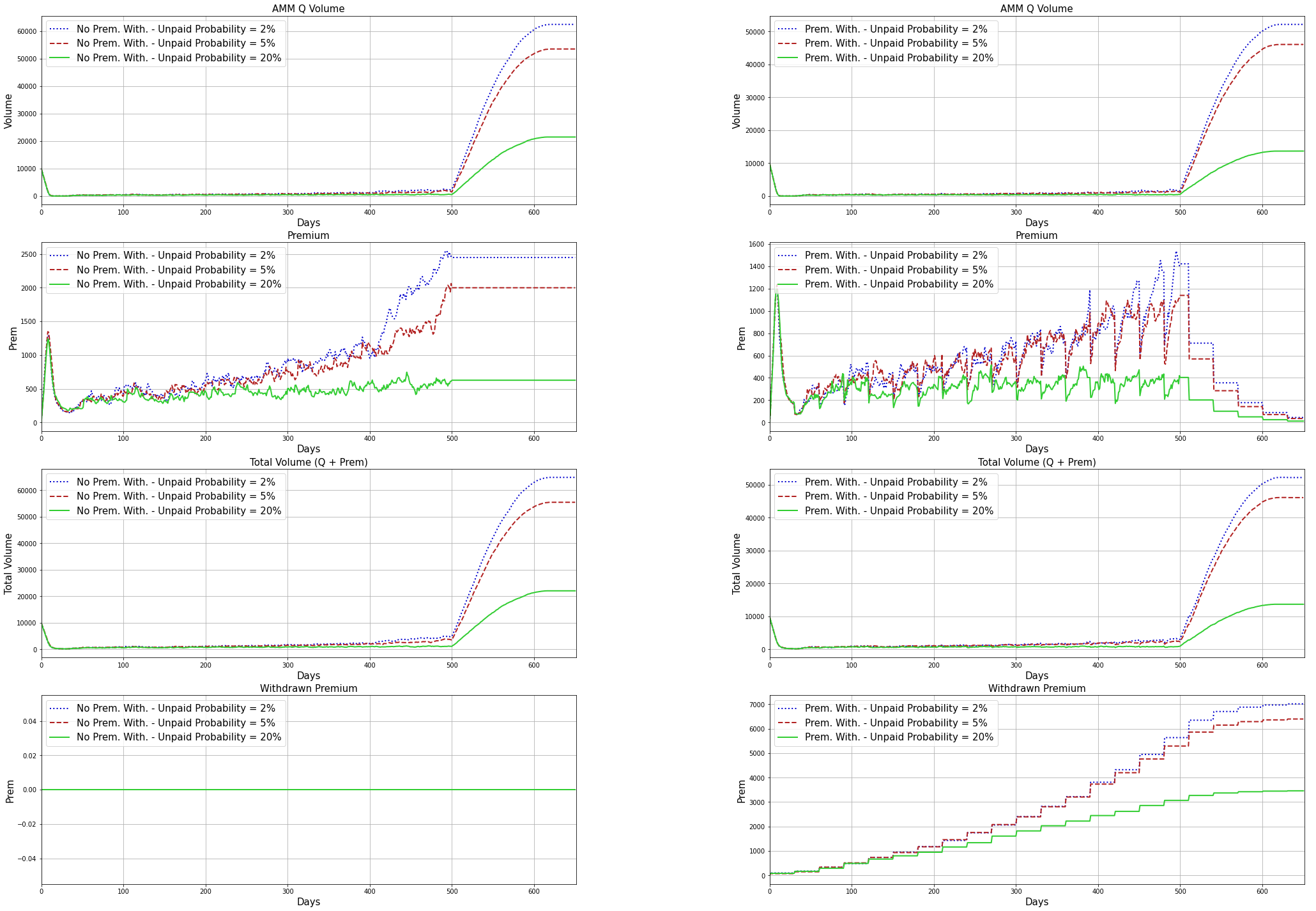}
        \caption{AMM Curves from simulation of scenario 2 and 30 days of withdrawal period.
        \label{scenario_2_all_curves_30days}}
\end{figure}
        \vfill

\end{landscape}

\begin{landscape}

\subsection[\appendixname~\thesubsection]{Scenario 2 - 90 days premium withdrawn}
\vfill
\begin{figure}[H]
        \centering
        \includegraphics[scale=0.35]{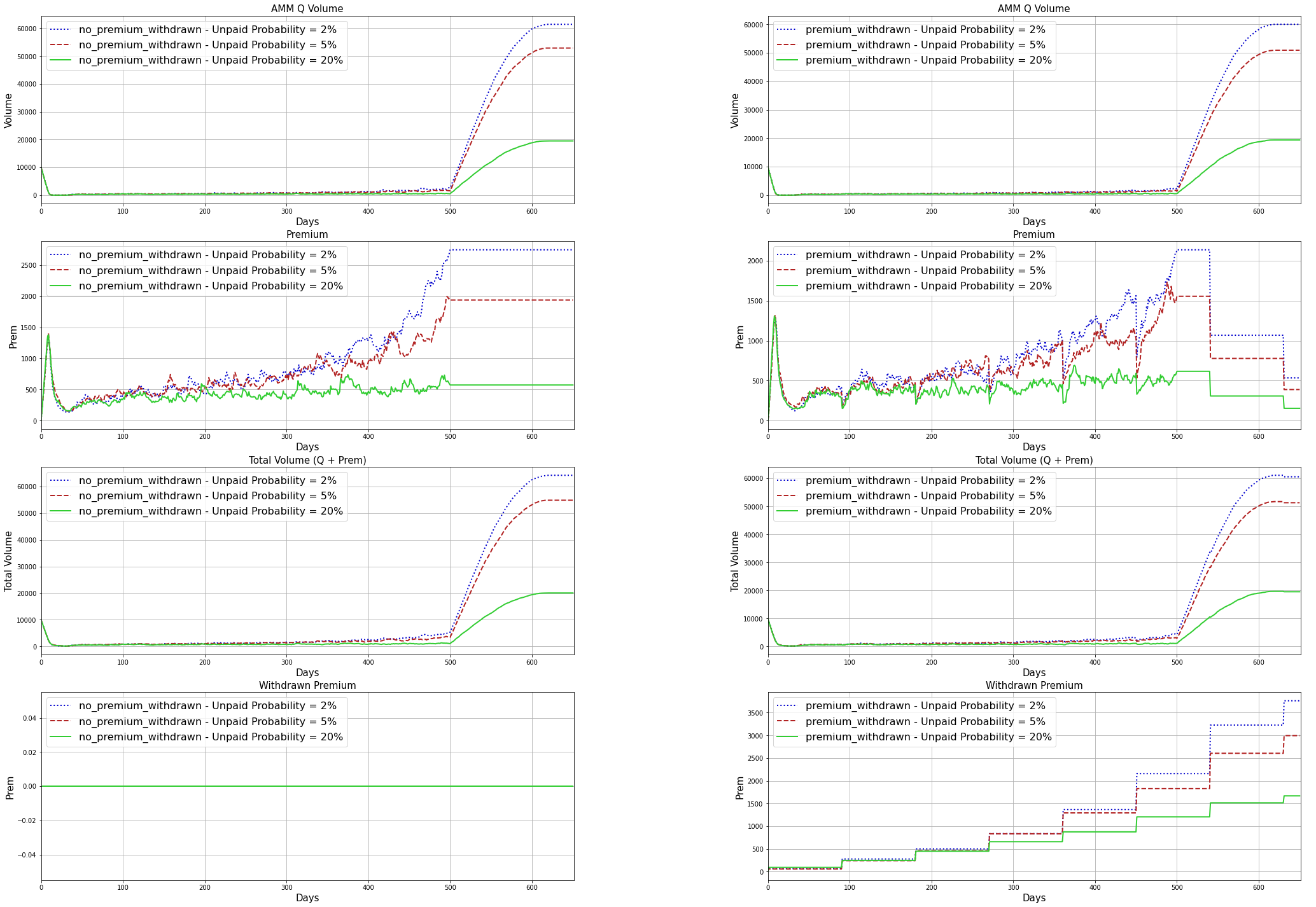}
        \caption{AMM Curves from simulation of scenario 2 and 90 days of withdrawal period.
        \label{scenario_2_all_curves_90days}}
\end{figure}
\vfill

\end{landscape}

\begin{landscape}

\subsection[\appendixname~\thesubsection]{Scenario 3 - 1 day premium withdrawn}
\vfill
\begin{figure}[H]
        \centering
        \includegraphics[scale=0.35]{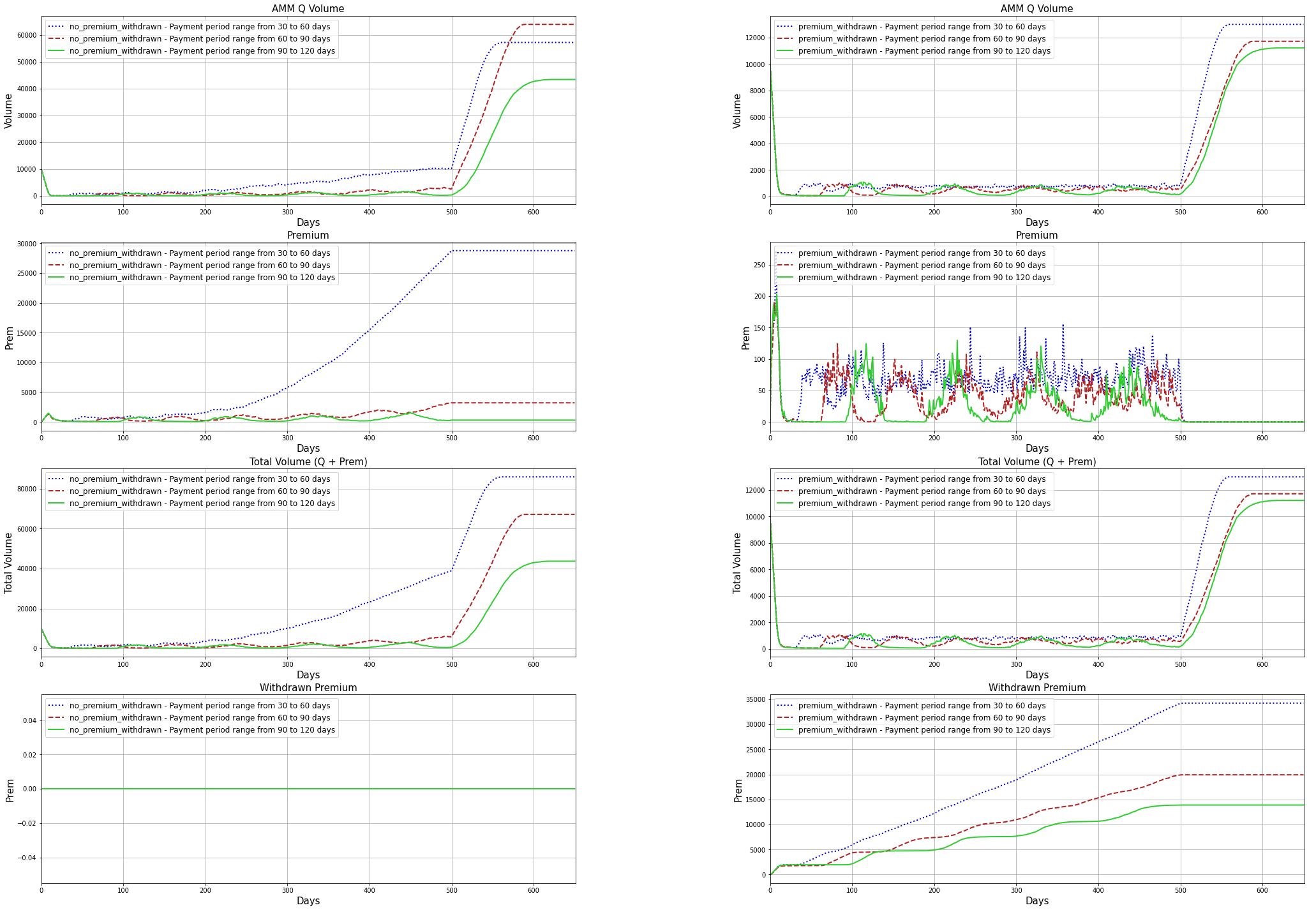}
        \caption{AMM Curves from simulation of scenario 3 and 1 day of withdrawal period.
        \label{scenario_3_all_curves_1day}}
\end{figure}
        \vfill

\end{landscape}

\begin{landscape}

\subsection[\appendixname~\thesubsection]{Scenario 3 - 30 days premium withdrawn}
\vfill
\begin{figure}[H]
        \centering
        \includegraphics[scale=0.35]{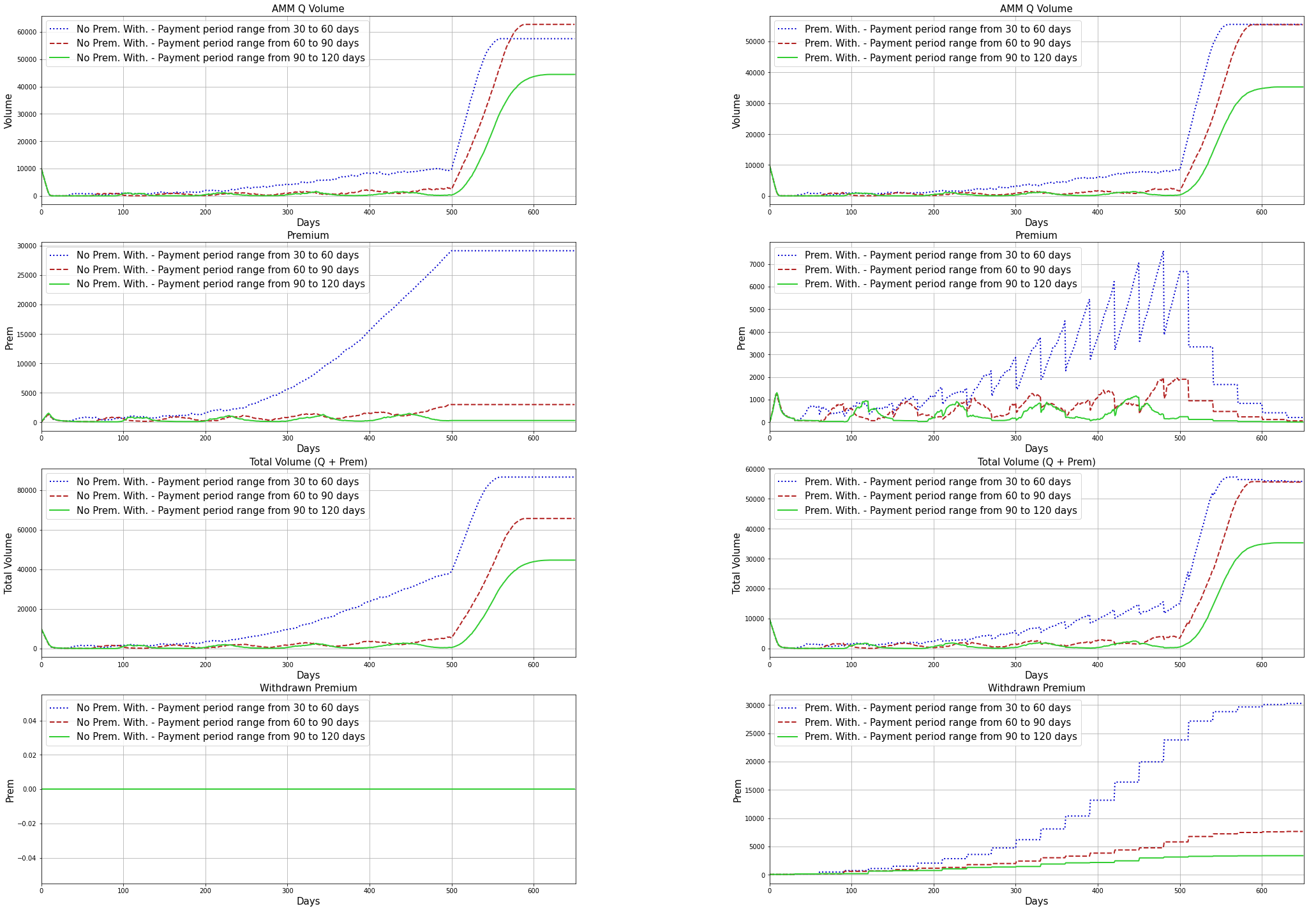}
        \caption{AMM Curves from simulation of scenario 3 and 30 days of withdrawal period.
        \label{scenario_3_all_curves_30days}}
\end{figure}
        \vfill

\end{landscape}

\begin{landscape}

\subsection[\appendixname~\thesubsection]{Scenario 3 - 90 days premium withdrawn}
\vfill
\begin{figure}[H]
        \centering
        \includegraphics[scale=0.35]{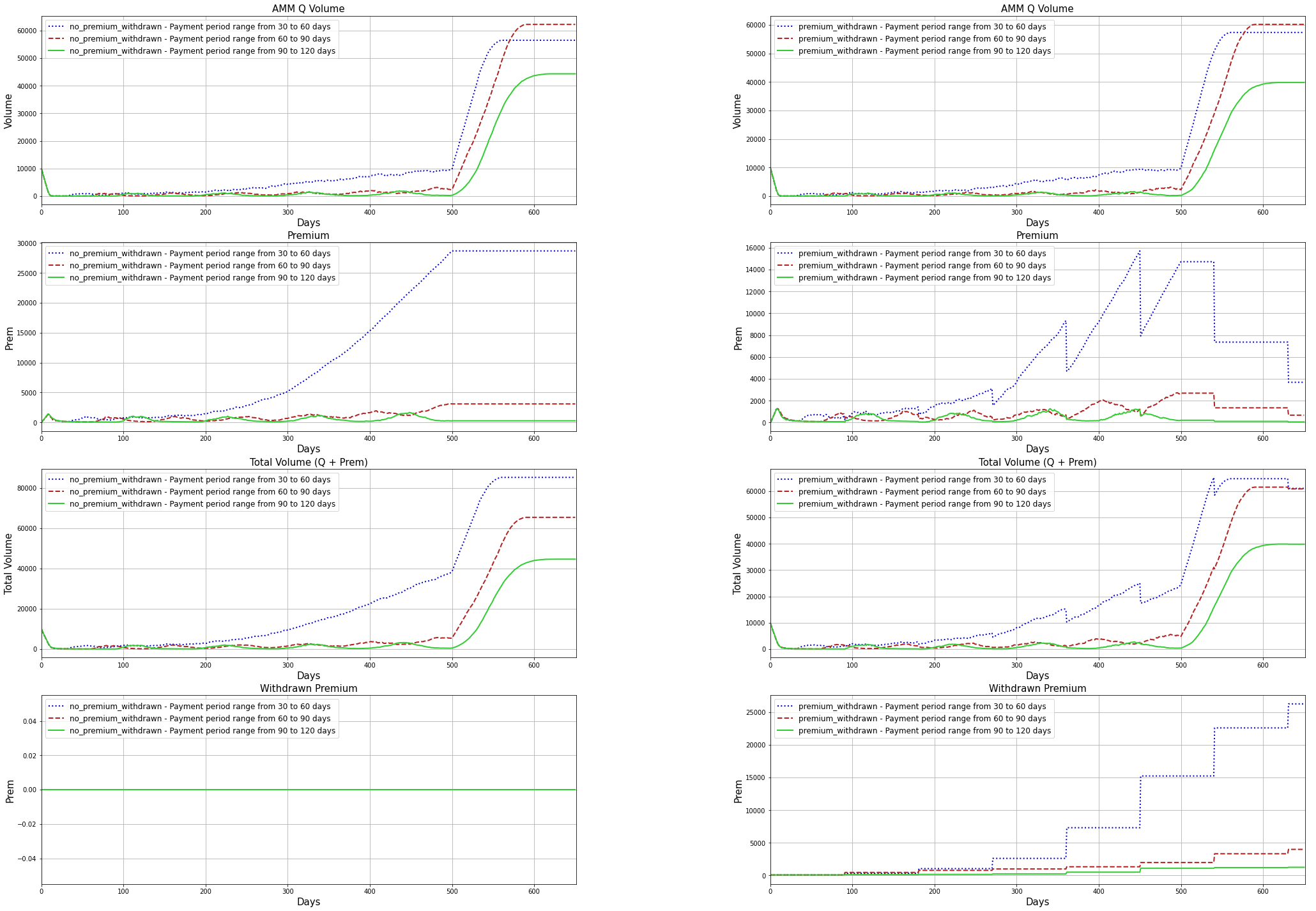}
        \caption{AMM Curves from simulation of scenario 3 and 90 days of withdrawal period.
        \label{scenario_3_all_curves_90days}}
\end{figure}
\vfill

\end{landscape}

\begin{landscape}

\subsection[\appendixname~\thesubsection]{Scenario 4 - 1 day premium withdrawn}
\vfill
\begin{figure}[H]
        \centering
        \includegraphics[scale=0.35]{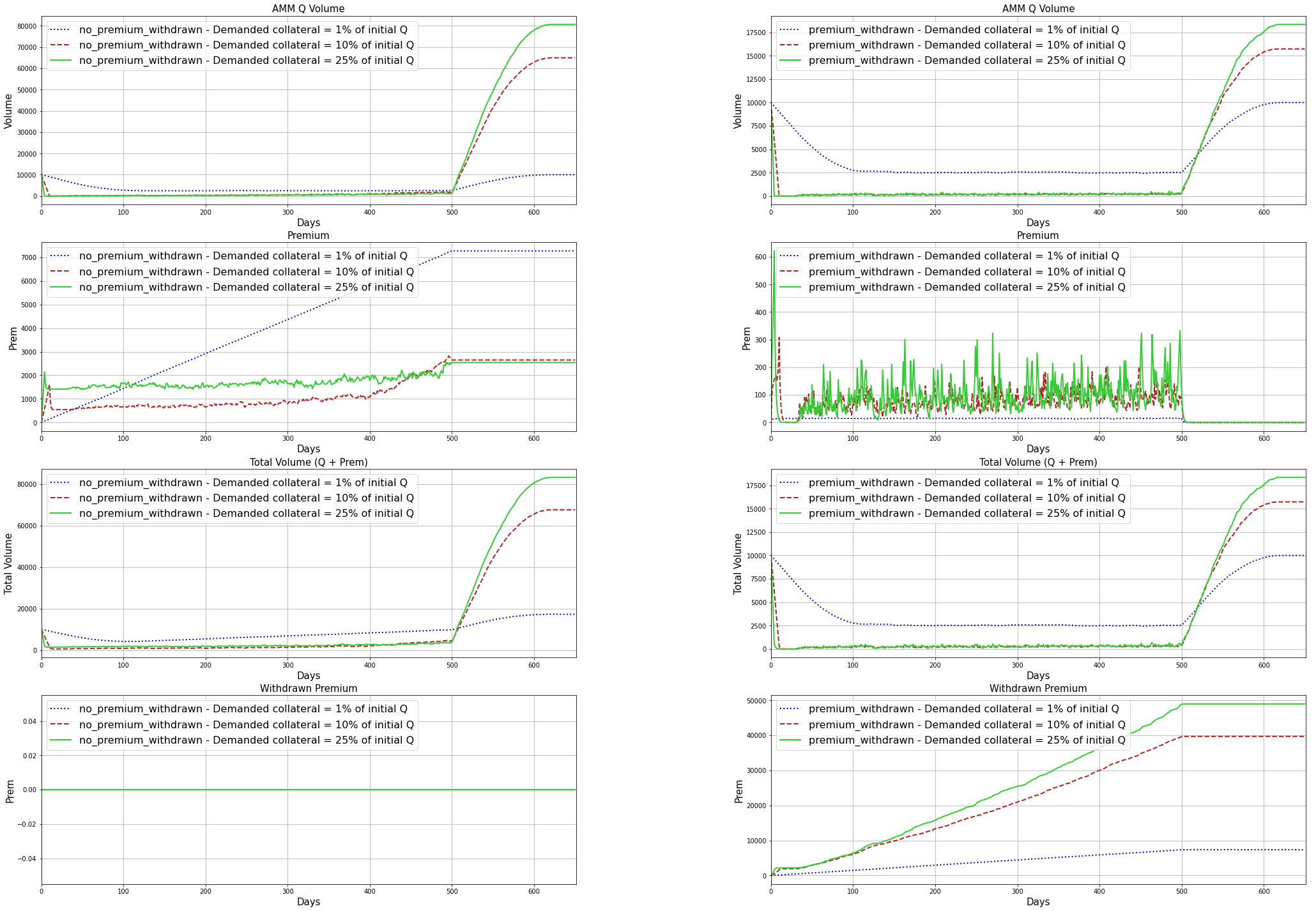}
        \caption{AMM Curves from simulation of scenario 4 and 1 day of withdrawal period.
        \label{scenario_4_all_curves_1day}}
\end{figure}
        \vfill

\end{landscape}

\begin{landscape}

\subsection[\appendixname~\thesubsection]{Scenario 4 - 30 days premium withdrawn}
\vfill
\begin{figure}[H]
        \centering
        \includegraphics[scale=0.35]{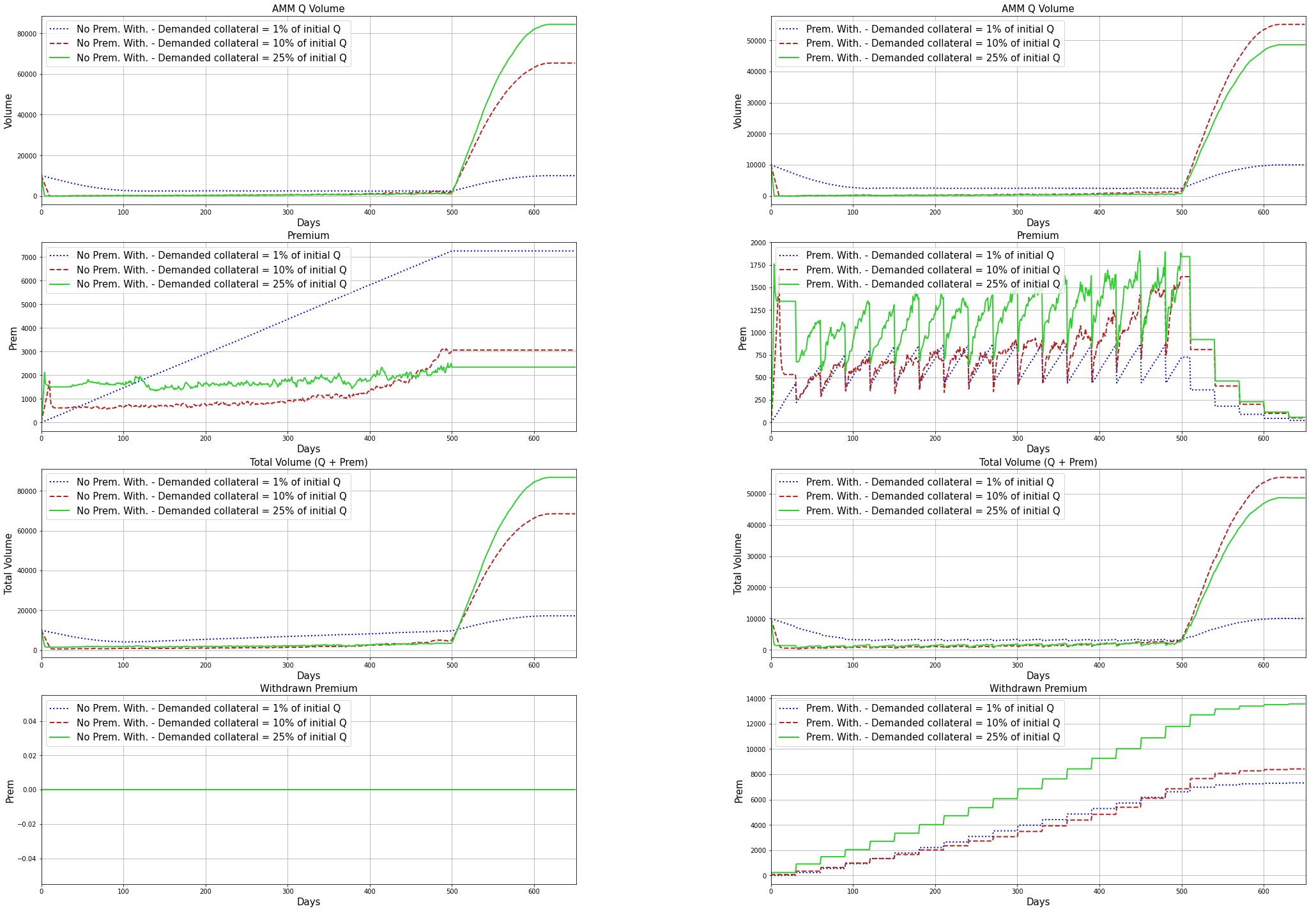}
        \caption{AMM Curves from simulation of scenario 4 and 30 days of withdrawal period.
        \label{scenario_4_all_curves_30days}}
\end{figure}
        \vfill

\end{landscape}

\begin{landscape}

\subsection[\appendixname~\thesubsection]{Scenario 4 - 90 days premium withdrawn}
\vfill
\begin{figure}[H]
        \centering
        \includegraphics[scale=0.35]{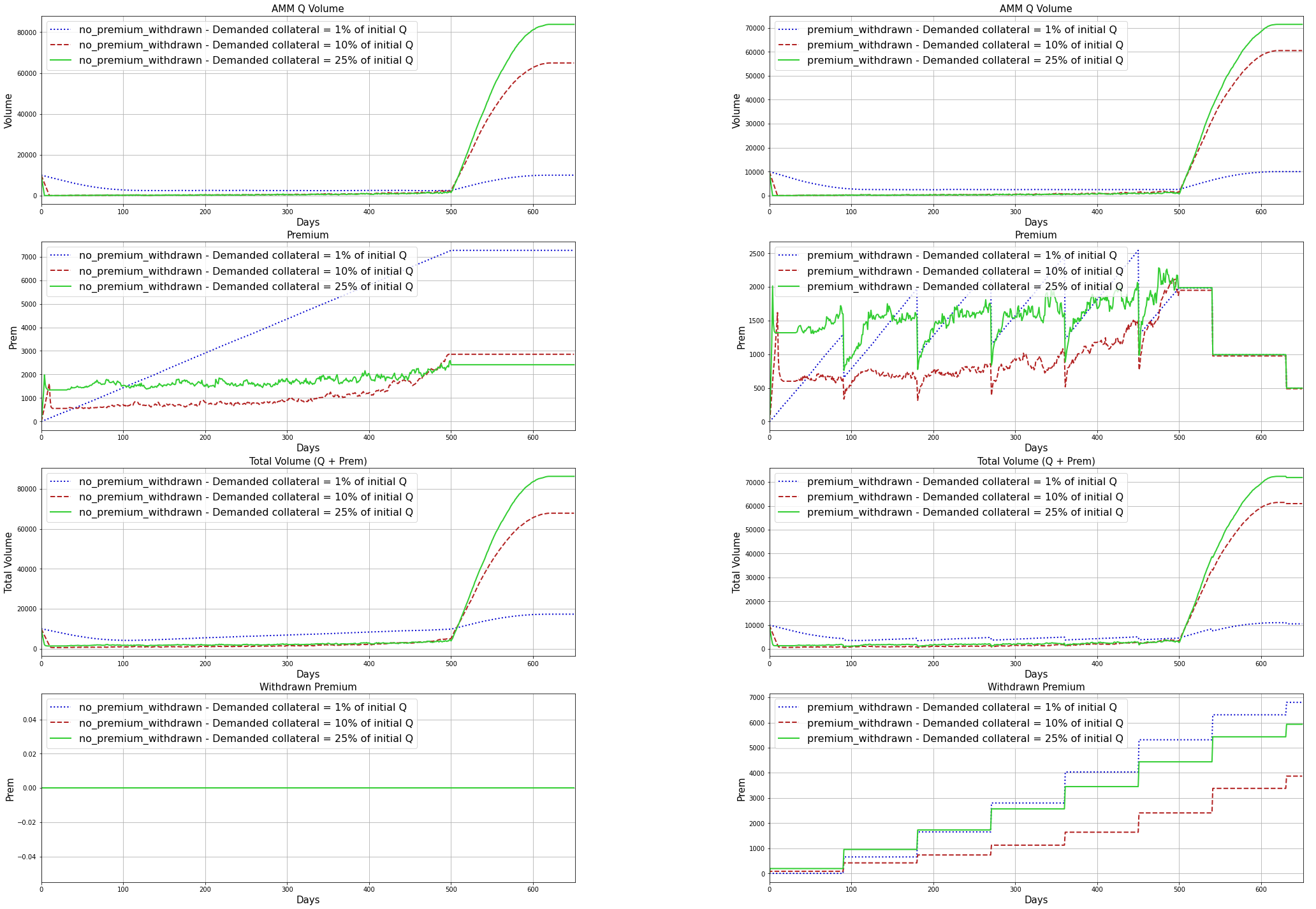}
        \caption{AMM Curves from simulation of scenario 4 and 90 days of withdrawal period.
        \label{scenario_4_all_curves_90days}}
\end{figure}
\vfill

\end{landscape}

\begin{landscape}

\subsection[\appendixname~\thesubsection]{Scenario 5 - 1 day premium withdrawn}
\vfill
\begin{figure}[H]
        \centering
        \includegraphics[scale=0.35]{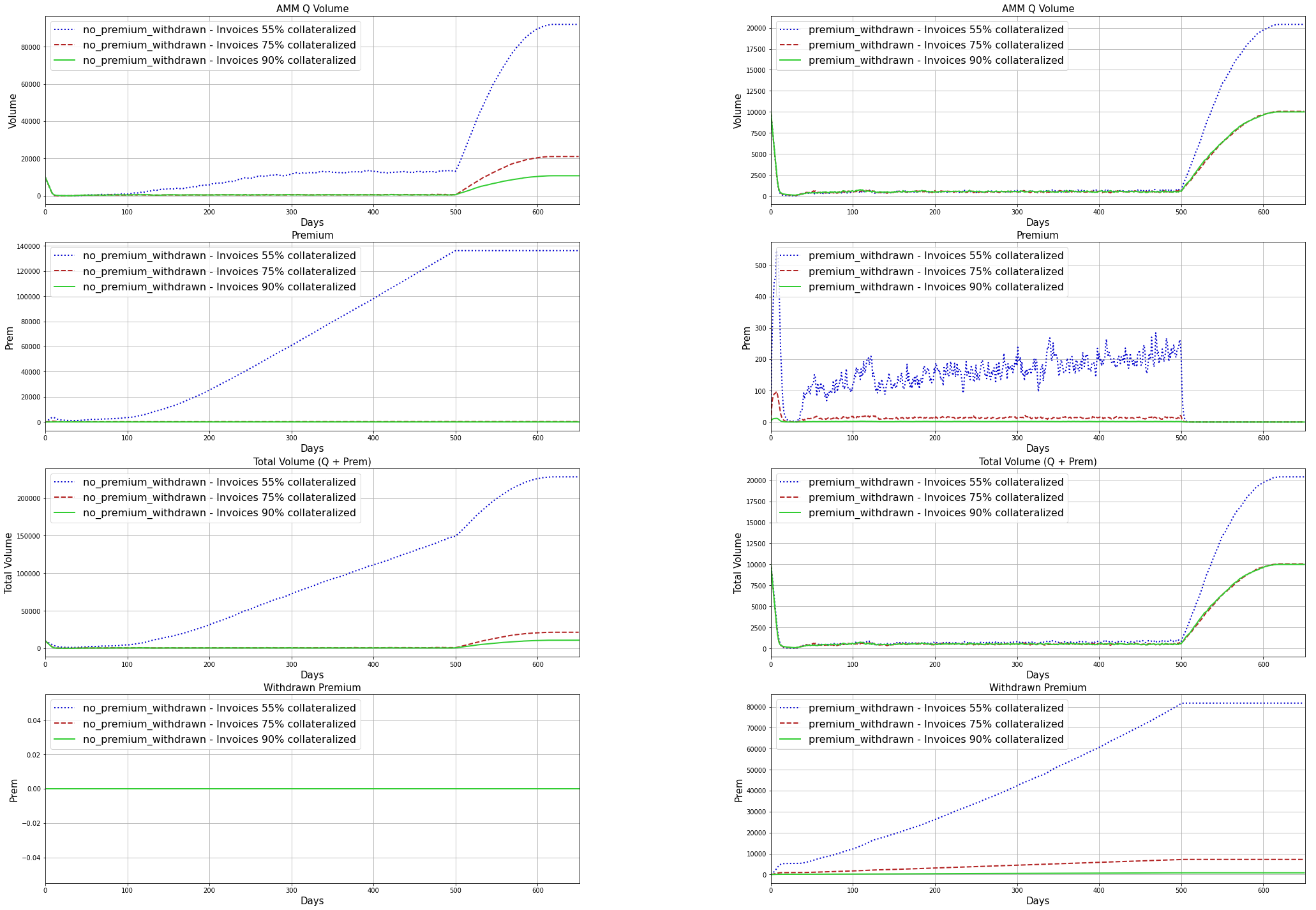}
        \caption{AMM Curves from simulation of scenario 5 and 1 day of withdrawal period.
        \label{scenario_5_all_curves_1day}}
\end{figure}
\vfill

\end{landscape}

\begin{landscape}

\subsection[\appendixname~\thesubsection]{Scenario 5 - 90 days premium withdrawn}
\vfill
\begin{figure}[H]
        \centering
        \includegraphics[scale=0.35]{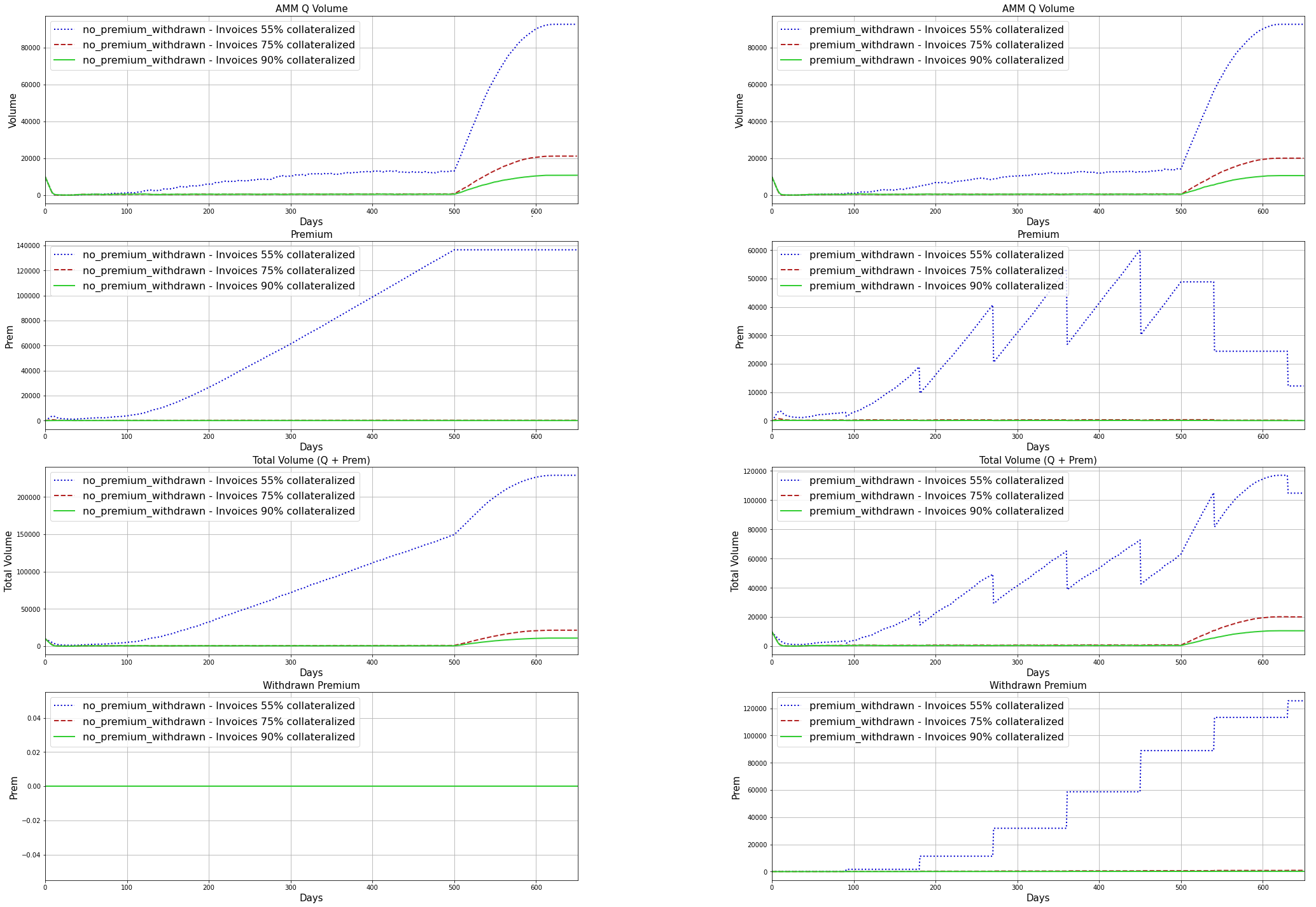}
        \caption{AMM Curves from simulation of scenario 5 and 90 days of withdrawal period.
        \label{scenario_5_all_curves_90days}}
\end{figure}
\vfill

\end{landscape}

\begin{landscape}

\subsection[\appendixname~\thesubsection]{Hack scenario - 1 day premium withdrawn}
\vfill
\begin{figure}[H]
        \centering
        \includegraphics[scale=0.35]{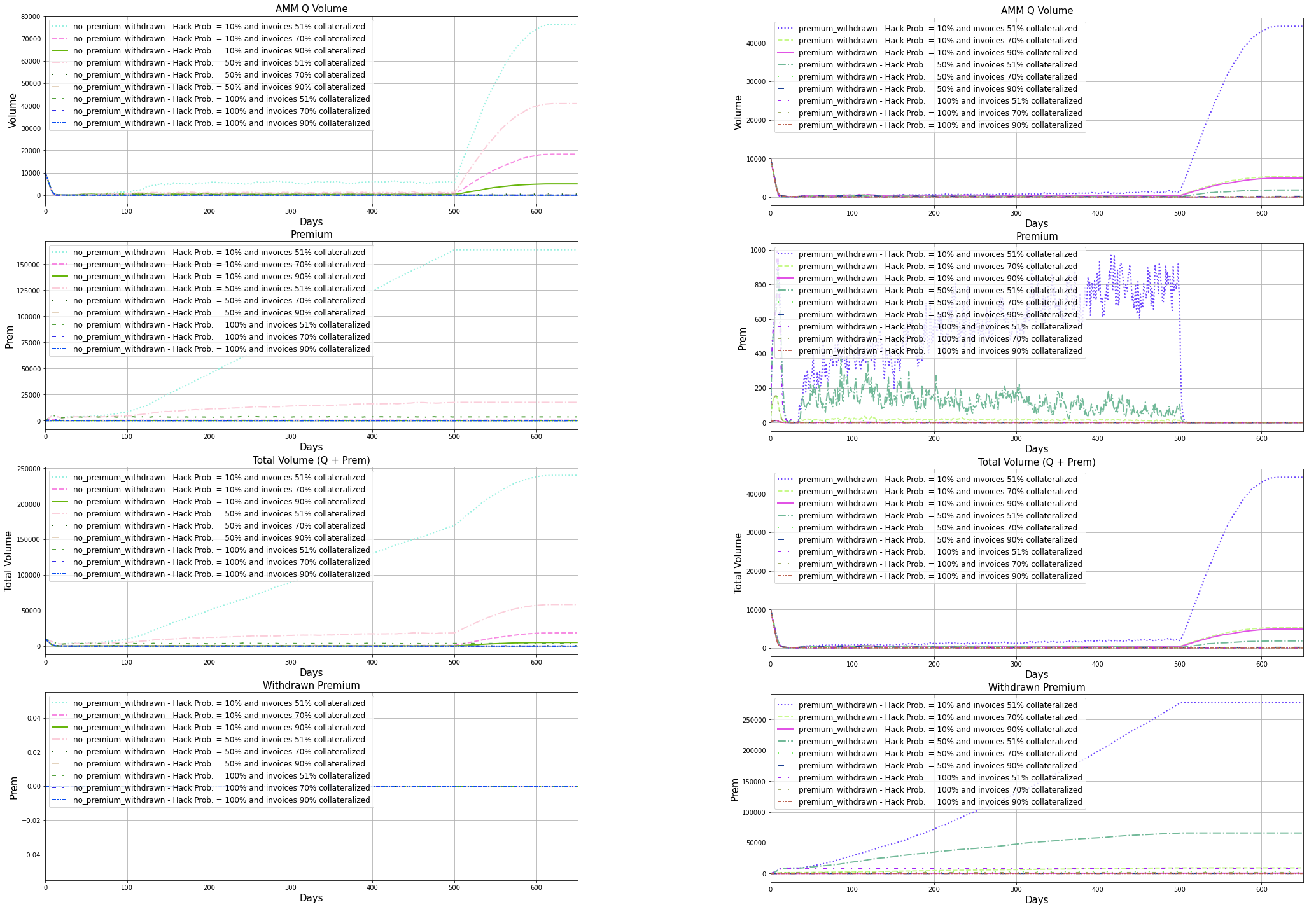}
        \caption{AMM Curves from simulation of hack scenario and 1 day of withdrawal period.
        \label{hack_scenario_all_curves_1day}}
\end{figure}
        \vfill

\end{landscape}

\begin{landscape}

\subsection[\appendixname~\thesubsection]{Hack scenario - 90 days premium withdrawn}
\vfill
\begin{figure}[H]
        \centering
        \includegraphics[scale=0.35]{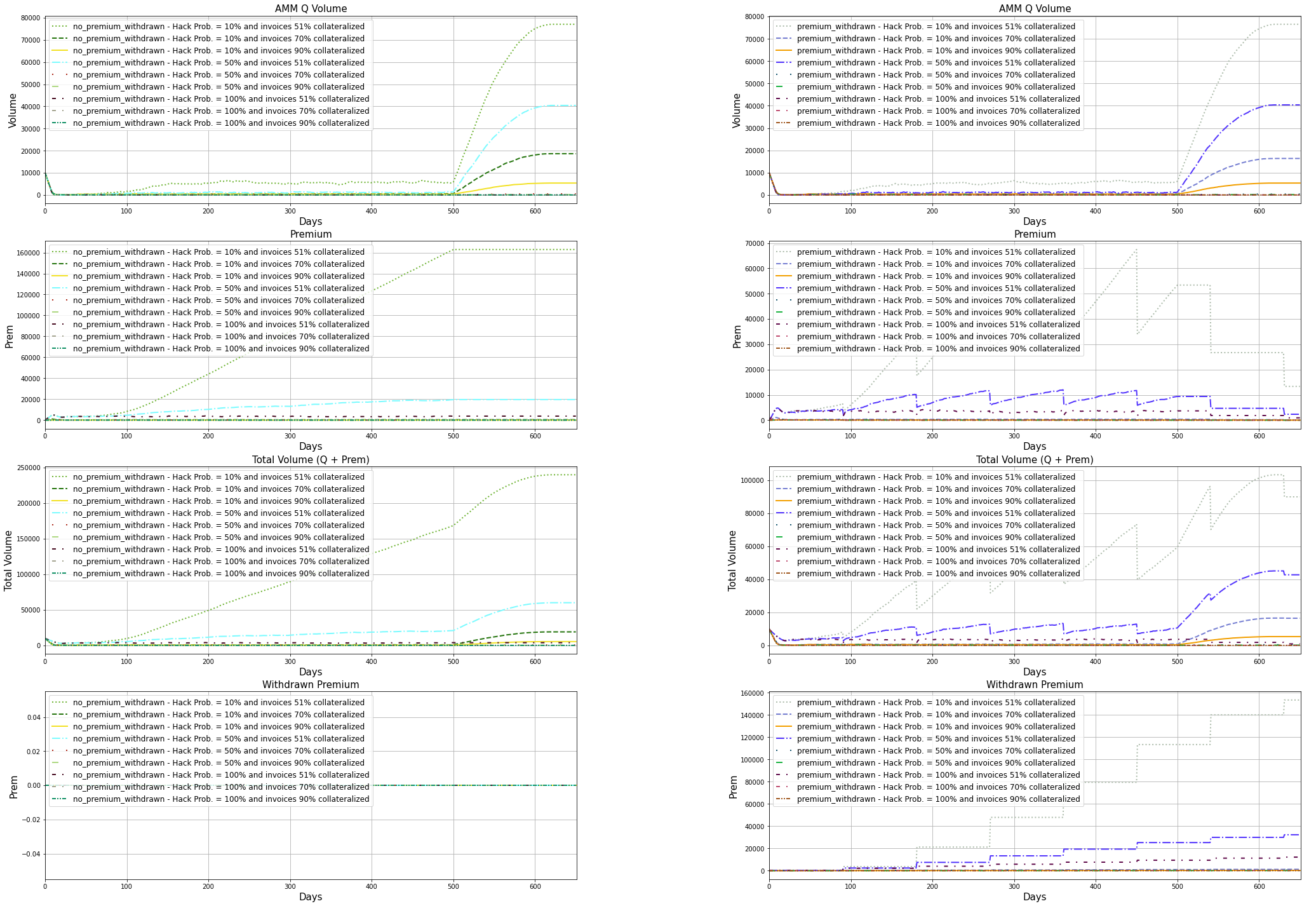}
        \caption{AMM Curves from simulation of hack scenario and 90 days of withdrawal period.
        \label{hack_scenario_all_curves_90days}}
\end{figure}
\vfill

\end{landscape}

\restoregeometry




\end{document}